\documentstyle[preprint,epsfig,aps]{revtex}
\tightenlines
\begin{document}
\draft
\preprint{\vbox{\hbox{IFIC/00-27}
 \hbox{hep-ph/0005244}}}
\title{MSW Solutions to the Solar Neutrino Problem in Presence
of Noisy Matter Density Fluctuations}

\author{A.A. Bykov$^1$, M.C. Gonzalez-Garcia$^2$, C. Pe\~na-Garay$^2$, 
V.Yu. Popov$^1$,V.B. Semikoz$^3$}

\address{$^1$ Department of Physics, Moscow State University\\
           119899, Moscow, Russia\\
	 $^2$ Instituto de F\'{\i}sica Corpuscular\\
              Universidad de Valencia-CSIC\\
              Edificio de Institutos de Investigacion, Apt 2085\\ 
	      46071 Val\`encia, Spain. \\
	 $^3$ The Institute of Terrestrial Magnetism, Ionosphere and \\
              Radio Wave Propagation of the Russian Academy of Sciences,\\
               IZMIRAN, Troitsk, Moscow region, 142190, Russia}

\maketitle
\begin{abstract}
We study the effect of random matter density fluctuations in the sun on 
resonant neutrino conversion in matter. We 
assume no specific mechanism for generation of the fluctuation and we keep 
the amplitude and correlation length as independent parameters. 
We do not work under the approximation that fluctuations have spatial 
correlations only over distances small compared to the neutrino oscillation 
lengths. Instead we solve numerically the evolution equation for
the neutrino system including the full effect of the random matter density 
fluctuations of given amplitude and correlation length. 
In order to establish the possible
effect on the MSW solutions to the solar neutrino problem we perform a global 
analysis of all the existing observables including the measured
total rates as well as the Super--Kamiokande measurement on the time 
dependence of the event rates during  the day and night and the 
recoil electron energy  spectrum. We find the effects of random noise
to be larger for small mixing angles and they are mostly important 
for correlation lengths  in the range  
few 100 km $\lesssim L_0\lesssim$ few 1000 km. They can be understood
as due to a parametric resonance occuring when the
phase acquired by the oscillating neutrino state on one fluctuation length 
$L_0$ is a multiple of 2$\pi$. 
We find that this resonant parametric condition is mainly 
achieved for low energy neutrinos such as the pp-neutrinos 
and therefore its effect is mostly seen on the 
total event rates while the other Super--Kamiokande observables
are very marginally sensitive to the presence of noise due to the
higher energy threshold. 
\end{abstract}
\newpage
\section{Introduction}
Solar neutrinos were first detected already three decades ago  
in the Homestake experiment \cite{homestake0}  
and from the very beginning it was pointed out the puzzling issue of  
the deficit in the observed rate as compared to the theoretical  
expectation based on the standard solar model \cite{SSMold}  
with the implicit assumption that neutrinos created in 
the solar interior reach the Earth unchanged, i.e. they are massless 
and have only standard properties and interactions.  
This discrepancy led to a change in the 
original goal of using solar neutrinos to probe the properties of the 
solar interior towards the study of the properties of the neutrino 
itself and it triggered an intense activity both theoretical as well 
as experimental, with new measurements being proposed in order to 
address the origin of the deficit.  
 
On the theoretical side, enormous progress has been done in the  
improvement of solar modelling and 
calculation of nuclear cross sections. For example,  
helioseismological observations 
have now established that diffusion is occurring and by now most solar 
models incorporate the effects of helium and heavy element 
diffusion~\cite{Bahcall:1997qw,Bahcall:1995bt}. 
From the experimental point of view the situation is now much 
richer. Four additional experiments to the original Chlorine experiment at  
Homestake \cite{homestake} have also detected solar neutrinos: 
the radiochemical Gallium experiments on $pp$ neutrinos, GALLEX 
\cite{gallex} and SAGE~\cite{sage}, and the water Cerenkov  
detectors Kamiokande~\cite{kamioka} and Super--Kamiokande 
\cite{sk1,sk99}. The latter have been able not only to confirm the original  
detection of solar neutrinos at lower rates than predicted by standard 
solar models, but also to demonstrate directly that the neutrinos come 
from the Sun by showing that recoil electrons are scattered in the 
direction along the Sun--Earth axis. Moreover, they have also provided 
us with useful information on the time dependence of the event rates during  
the day and night, as well as a measurement of the recoil electron energy  
spectrum. After 825 days of operation, Super--Kamiokande has also presented 
preliminary results on the seasonal variation of the neutrino event 
rates, an issue which will become important in discriminating the MSW 
scenario from the possibility of neutrino oscillations in 
vacuum~\cite{ourseasonal,v99}. At the present stage, the quality 
of the experiments themselves and the  
robustness of the theory give us confidence that in order to  
describe the data one must depart from the Standard Model (SM) of particle  
physics interactions by  
endowing neutrinos with new properties. In theories beyond the SM,  
neutrinos may naturally have new properties, 
the most generic of which is the existence of mass.   
It is undeniable that the most popular explanation of the solar  
neutrino anomaly is in terms of neutrino masses and mixing leading to neutrino 
oscillations either in {\sl vacuum}~\cite{Glashow:1987jj} or via the 
matter-enhanced {\sl MSW mechanism}~\cite{msw}. 
 
The standard MSW analysis is based on a mean--field treatment of the
solar background through which the neutrinos propagate. In this approximation
the global analysis of the full neutrino data sample 
described above \cite{oursolar} leads to the existence of three 
allowed regions in the $\Delta m^2$ $\sin^2 2\theta$ parameter space
for neutrino oscillations
\begin{itemize}
\item[$\bullet$] non-adiabatic-matter-enhanced oscillations 
or small mixing angle (SMA) region  
with $ \Delta m^2=(0.4$--$1)\times 10^{-5}$ eV$^2$ and 
$\sin^2(2\theta)=(1$--$10)\times 10^{-3} $, and 
\item[$\bullet$] large mixing (LMA) region  
$\Delta m^2=(0.2$--$5)\times 10^{-4}$  eV$^2$ and 
$\sin^2(2\theta)=0.6$--$1$.
\item[$\bullet$] low mass solution (LOW) 
$\Delta m^2=(0.3$--$2)\times 10^{-7}$  eV$^2$ and 
$\sin^2(2\theta)=0.8$--$1$.
\end{itemize}

There are several works in the literature 
\cite{Krastev,otherper,other,Burgess0,Nunokawa,Burgess} 
where corrections to such mean--field picture have been studied. 
The influence of periodic matter density fluctuations of given 
amplitude and fixed frequency  
above the average density on resonant neutrino conversion 
was investigated in Refs.\cite{Krastev,otherper}. In Ref.~\cite{Krastev}
a parametric resonance is found when the fixed frequency of the perturbation
is close to the neutrino oscillation frequency. This approach however
gives not answer to the physical origin of such fixed frequency perturbation.

More recently, the main approach to fluctuations that has been pursued 
\cite{other,Burgess0,Nunokawa} is to model the matter density as a Gaussian 
random variable (white noise). In this approach the number of free parameters 
remains the same as in the case of fixed frequency perturbations -- 
two, the perturbation amplitude and the correlation length-- 
but the presence of white noise in the sun is doubtless since there are 
many mechanism to generate random density perturbations. For technical reasons 
these analysis where performed  under the assumption that fluctuations have 
spatial correlations only over distances small compared to the neutrino 
oscillation lengths. Within this approximation the conclusions obtained 
were that such fluctuations in the solar electron density can significantly 
modify the MSW solutions to the solar neutrino problem (SNP)
provided that their relative amplitude near the MSW resonance point can be as
large as few percent. 

In Ref.~\cite{Burgess} criticisms to these results
were raised based on two facts: i) the unexistence of a plausible source for
such $\delta$-correlated fluctuations in the vicinity of the MSW resonance
point and ii) whatever its origin, the effect of the density perturbation
is maximum in the regime where the short correlation--length approximation
fails. In particular in Ref.~\cite{Burgess} they concentrate on 
helioseismological waves as origin of the perturbation and in particular
on g-waves whose amplitude increases with the solar depth and can, 
in principle, reach the interesting values to affect neutrino propagation. 
However they conclude that such g-waves do not affect the MSW neutrino 
conversions since the wavelength for the lower modes, for which the largest 
amplitudes are possible, 
is much longer than the characteristic MSW neutrino oscillation length.

It has been recently argued \cite{Dzhalilov}, however, that in the 
magnetohydrodynamical (MHD) generalization of the Helioseismology 
the objection in Ref.~\cite{Burgess} does not hold.
Assuming modest central large-scale magnetic fields ($B_0=$1--10--100~Gauss)
one can find magneto-gravity eigenmodes with much shorter wave lengths for 
density perturbations $\lambda_{MHD}\sim 200-2000~$km
\cite{Dzhalilov} that is comparable with the neutrino oscillation length 
at the MSW resonance for large and small mixing angles correspondingly.

In this paper we revisit the problem of the effect of matter of density  
fluctuations in the sun on the MSW solutions to the SNP. In our approach we 
assume no specific mechanism for generation of the fluctuation and we keep 
the amplitude and correlation length as independent parameters. There are 
two main differences in our analysis as compared to those in Refs.~
\cite{other,Burgess0,Nunokawa,Burgess}. 
First we do not work under the approximation that fluctuations have spatial 
correlations only over distances small compared to the neutrino oscillation 
lengths. Instead we solve numerically the evolution equation for
the neutrino system including the full effect of the random matter density 
fluctuations of given amplitude and correlation length. 
Second, in order to establish the possible
effect on the MSW solutions to the SNP we perform a global analysis of all 
the existing observables including not only the measured total rates 
but also the Super--Kamiokande measurement on the time dependence of the 
event rates during  the day and night, as well as the recoil electron energy  
spectrum. In particular we include the regeneration effects when neutrinos
cross the Earth \cite{daynight} which were neglected in Ref. \cite{Nunokawa}.

The outline of the paper is as follows. In Sec.~\ref{solutions} 
we discuss our approach to the solution of the neutrino evolution
in the presence of density fluctuations. In Sec.~\ref{parke} we 
briefly summarize the standard analytical approach based on the short 
correlation length approximation and in Sec.~\ref{numerical} we
discuss our numerical treatment and present our results for the
survival probabilities as a function of the relevant oscillation
and noise parameters. In Sec.~\ref{resonance} we interpret our  
results for the enhancement of the survival probability in the 
language of parametric resonance in MSW conversions. 
Section~\ref{analysis} is devoted to the statistical analysis
of the solar neutrino observables in the framework of the
MSW solutions of the SNP in presence of the noisy density 
fluctuations. We study the variation of the allowed regions of the SNP for 
different combinations of observables when noise fluctuations with  
different correlation lengths are included.
Our results are summarized in Figs.~\ref{rates}--\ref{global}
and Table~\ref{chimin}. We show that even for noise levels as large
as 4\% the relative quality of
the three allowed regions of the MSW solutions to the SNP depends on the 
value of the correlation length studied in the wide range $L_0=70$--$10^4$ 
Km and that the three allowed regions of the MSW solutions of the SNP
remain valid at the present level of solar neutrino experiments.
Finally in Sec.~\ref{conclu} we discuss our results and summarize our 
conclusions.
 
\section{MSW solutions to SNP for noisy matter density in the Sun.}
\label{solutions}
Some mechanisms to rise density perturbations in the Sun were 
proposed in literature in connection with the MSW solution to the SNP.
One of them with g-waves in Helioseismology as a plausible source of matter
noise fails being applied to the MSW neutrino conversions since the 
wavelength of such modes happens to be too long, $\lambda_g\sim 0.1R_{\odot}\gg
l_{osc}$ to influence neutrino oscillations\cite{Burgess}.
This $\lambda_g$ is the wave length for low
radial degree $n \leq 3$ for which largest g-mode amplitudes 
$\delta \rho (r)/\rho_0(r)\sim 4$ \% are possible \cite{Burgess}.

However, in the MHD generalization of the Helioseismology
such objection\cite{Burgess} does not apply. 
Assuming modest central large-scale magnetic fields ($B_0= 1-10-100$ Gauss)
one can find magneto-gravity eigenmodes with much shorter wave lengths for 
density perturbations $\lambda_{MHD}\sim l_{osc}\sim 200-2000$ km  
\cite{Dzhalilov} comparable with the neutrino oscillation length 
at the MSW resonance for large and small mixing angles correspondingly.
  
Note that standard Helioseismology corrections to the standard solar
model (SSM) (neglecting magnetic 
fields in the Sun) give density fluctuations deep in solar interior at a 
low level $\delta \rho (r)/\rho_0(r)\lesssim 1$\% \cite{Dziembowski}. 
This analysis is done using the solution of the inverted problem in 
Helioseismology based on the corresponding integral equation for which 
the kernels are built on the full set of p-mode waves calculated and 
observed on the photosphere
\cite{Dziembowski1}. However, such analysis does not touch both g-modes 
that are still invisible on the surface of the Sun (from SOHO satellite)
and has no relation to MHD modes found in \cite{Dzhalilov}.
%
\subsection{Generalized Parke formula for averaged evolution equation}
\label{parke}
In this subsection we show that the Schr\"{o}dinger equation approach for 
noisy matter \cite{Nunokawa} generalized here for an arbitrary matter density 
perturbation correlator (not only $\delta$-correlator like in \cite{Nunokawa})
is equivalent to the Redfield evolution equation leading to the 
generalized Parke formula for the survival probability in the presence of noise
\cite{Burgess}.

Let us assume the presence of regular density 
waves $Re~\Bigl ({\displaystyle \sum_n} C_n\delta \rho_n(t)\Bigr )$
exited somehow within the solar interior. They could be,
for instance, the MHD matter density waves, $\delta \rho_n(z)$, 
which appear in the 1-dimensional solar model with 
the exponential matter background profile $\rho_0(z)\sim \exp (-z/H)$, 
$H \simeq 0.1R_{\odot}$, in the presence of gravity {\bf g} = (0,0, -g) 
and an external constant magnetic field {\bf B} = $(B_0,0,0)$. 
Such waves obey the dispersion relation $\omega_n = \omega (n, B_0, k_x, k_y)$ 
with the periods $T = 2\pi/\omega_n\sim$ few days and they are quite different 
from the g-modes in Helioseismology. In particular, unlike
the helioseismological g-modes, they have very short 
wavelength along the z-axis $\lambda_z \sim R_{\odot}/n \ll \lambda_g$ 
for large node numbers $n\gg 1$ acceptable in the model \cite{Dzhalilov}.

Assuming these density perturbations added to the SSM background density 
profile $\rho_0(t)$, we write the master  Schr\"{o}dinger equation for MSW 
conversions of two neutrino flavors, $\nu_e\to \nu_y$,
\begin{equation} 
i\left( 
\begin{array}{l}       
\dot{\nu}_{e}\\ 
\dot{\nu}_{y} 
\end{array} 
\right) = 
\left( 
\begin{array}{cc} 
H_e &s_2\delta \\ 
s_2\delta & 0
\end{array} 
\right) 
\left( 
\begin{array}{c} 
\nu_{e}\\ 
\nu_{y} 
\end{array} 
\right)~, 
\label{master} 
\end{equation} 
where in the diagonal entry 
$H_e = V_{ey}(t)[1 + Re~\Bigl ({\displaystyle \sum_n} C_n\phi_n(t)\Bigr )]
- 2c_2\delta$  any density eigenmode  $\xi_n(t)=C_n\phi_n(t) = 
\delta \rho_n(t)/\rho_0(t)$ has the small amplitude, $\mid \xi_n(t)\mid\ll 1$.
$c_2 = \cos 2\theta$, $s_2 = \sin 2\theta$ and $\delta = \Delta 
m^2/4E$ are the neutrino mixing parameters; $V_{ey}(t) = 
G_F\sqrt{2}(\rho (t)/m_p)(1 - Y_n)$ and $V_{ey}=V_{es}(t) = G_F\sqrt{2}
(\rho (t)/m_p)(1- 3Y_n/2)$ are the neutrino vector potentials in the Sun  
for active-active (y= x) and for active-sterile neutrino conversions 
correspondingly. They are given by the neutron abundance 
$Y_n = m_p N_n(t)/\rho (t)$ where for neutral matter the relation $Y_e = 
Y_p = 1 - Y_n$ has been used and the SSM density profile $\rho 
(t)$ is given by BP98 model. In what follows 
we will discuss only conversion into active neutrinos.

Using the survival probability $P_{ee}=\nu_e^*\nu_e$ and the auxiliary 
functions $I= Im~(\nu_e^*\nu_y)$, $R= Re~(\nu_e^*\nu_y)$ one can derive 
from the master  equation above the equivalent system of dynamical equations. 
After averaging of those dynamical equations over small density 
perturbations 
$\mid \xi_n(t)\mid\ll 1$, such system takes the form: 
\begin{equation} 
i\frac{d}{dt}\left( 
\begin{array}{l}       
\cal{R}\\ 
\cal{I}\\
\cal{P}- \mbox{1/2}
\end{array} 
\right) = 
\left( 
\begin{array}{ccc} 
-2\kappa (t) &-(V_{ey}(t) - 2\delta c_2)&0\\ 
 V_{ey}(t) - 2\delta c_2 & -2\kappa (t)&\ -\delta s_2\\
0&+\delta s_2&0
\end{array} 
\right) 
\left( 
\begin{array}{c} 
\cal{R}\\ 
\cal{I}\\
\cal{P} - \mbox{1/2} 
\end{array} 
\right)~. 
\label{master1} 
\end{equation} 
with $\cal{P}=<P_{ee}>$, $\cal{I} =<I>$, $\cal{R} = <R>$.
 This system of equations is similar to Eq. (3.14) in \cite{Nunokawa} but here the matter perturbation parameter $\kappa (t)$ is of the form:
\begin{equation} 
\kappa (t) = \frac{1}{2}\sum_{m,n}\int_{t_1}^{t} V_{ey}(t)V_{ey}
(t_2)\langle C_nC_m\rangle\phi_n(t)\phi_m(t_2)dt_2~,
\label{kappa}
\end{equation} 
where in averaging $\langle ...\rangle$ we use that: (i) $C_n$ are 
uncorrelated random 
variables which are Gaussian distributed with vanishing mean: 
$\langle C_n\rangle=0$, (ii)
different modes (with different node number) are uncorrelated, 
$\langle C_n C_m \rangle = A_n\delta_{nm}$. 

It is reasonable to assume that even for a regular eigenmode ``n'' 
excited somehow within the solar interior, 
different neutrinos emitted from different starting points $t_0$ in the core 
propagate to the Earth along parallel rays crossing different profile 
realizations of the same mode $\xi_n(t)$.  
The phase of the wave entering in $C_n$ is 
random, $\langle C_n\rangle =0$, as well as a starting point 
$t_0$ is random for given 
t). In other words, the averaging over random phases is equivalent to the 
averaging over production point distribution (see 
the discussion in Sec.~\ref{numerical}).

Thus, one can consider the sum of multimode {\it regular} perturbations as 
{\it random density perturbations} 
$\sum C_n\phi_n(t)\to \xi (t)$, 
$\langle \xi (t)\rangle =0$, for which the $\delta$-correlated noise 
\begin{displaymath}
\langle\xi (t_1)\xi (t_2)\rangle  = 
L_{0}\langle\xi (t_1)^2\rangle\delta (t_1 -t_2)
\end{displaymath}
is the particular case 
leading from Eq.~(\ref{kappa}) to 
$\kappa (t) = V_{ey}^2(t)\langle\xi^2\rangle L_0/4$ ~\cite{Nunokawa}.

We can now identify the entries in the Hamiltonian Eq.~(\ref{master1}) 
with the corresponding terms in Eq.~(77) derived in Ref.~\cite{Burgess0} 
through the averaging of the Redfield equation for the density matrix. 
One finds full coincidence with the notation in Ref.~\cite{Burgess0}: 
$\rho_1 =\cal{I}$, $\rho_2 = \cal{R}$, $\rho_3 = \cal{P} - \mbox{1/2}$ and 
$a(t)=\kappa(t)$, $-2(M_3 + b)= V_{ey}(t) - 
\delta c_2$, $M_2=0$, $M_1 = \delta s_2$. 

For slowly varying variables $V_{ey}(t)$ and $\kappa (t)$, one can diagonalize 
the Hamiltonian in Eq.~(\ref{master1}) obtaining the generalized Parke 
formula (Eq.~(86) in Ref.~\cite{Burgess0})
\begin{equation}
P_{ee}(t)=\frac{1}{2} + \left(\frac{1}{2}-P_J\right)
\exp{\left(-2\int_{t_0}^t \gamma_0(x) dx\right)} \cos 2\theta_m(t_0)\cos 2\theta_m(t)
\end{equation}
where the effect of the density fluctuations is contained in the
$\gamma_0$ factor 
\begin{equation}
\gamma_0(x)=\frac{4 \kappa(x) \delta^2 s_2^2}
{4 \delta^2 s_2^2+(V_{ey}(x)-\delta c_2)^2}\;. 
\label{gparke}
\end{equation} 
$P_J$ is the probability for level crossing as one passes the resonant 
point \cite{LZ} and $\theta_m$ is the mixing angle in matter.

\subsection{Computer simulation of the master equation}
\label{numerical}

There are some remarks to the analytic approach shown above that forced us 
to perform a numerical calculation. First, the generalized Parke formula 
in Eq.~(\ref{gparke}) does not describe neutrino conversions for large 
correlation lengths $L_0\gg l_{osc}$ in an 
appropriate way giving a huge discrepancy with the results from numerical 
calculations \cite{Burgess}. Alternatively, in 
Ref.~\cite{Burgess}, the survival probability was evaluated using the standard
Parke formula for each neutrino ray and then averaging this result over 200 
random density profiles of the Cell type of length $L_0$. 
Below we refer to this procedure as the ``Cell'' model.
Their result for large $L_0$, however, presents a strong dependence on the 
resonance position within a Cell. 

\begin{figure}
\begin{center}
\mbox{\epsfig{file=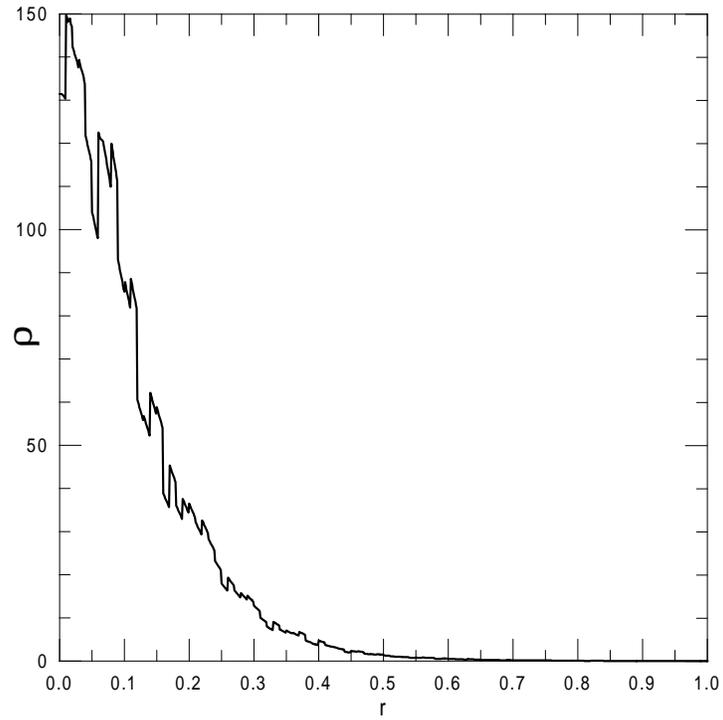,width=0.6\textwidth}}
\end{center}
\caption{Numerical density profile as a fuction of the radial distance 
for $L_0= 700$~km and noiselevel 
$\protect\sqrt{\langle\xi\rangle^2}= 0.1$.}
\label{rho} 
\end{figure}

In our approach we directly solve the Schr\"oedinger equation for the
two--neutrino system with the following procedure.
Not appealing to any origin of the matter density perturbations we 
consider different levels of the random density fluctuations 
$\langle\xi (r)\rangle\equiv\langle\delta \rho (r)/\rho_0(r)\rangle=0$
added to the background matter density $\rho_0(r)$ in SSM which we
take to be the BP98 density \cite{BP98}
\begin{equation}
\rho (r) =\rho_0(r) [1 + \xi (r)]~,
\label{noisedensity}
\end{equation}
where the parameter $\xi\equiv\sqrt{\langle\xi(r)^2\rangle}$ 
measures the amplitude of the perturbation. For a given 
value of $\xi$ and correlation length $L_0$ we generate 
a density profile of the type in Eq.(\ref{noisedensity}). The function 
$\xi(r)$ is constructed as steps of constant values 
$\xi\_i $ for each step $i$ of length $L_0$ from the production 
point to the edge of the Sun. The $\xi_i$ numbers are 
randomly generated following a Gaussian distribution of mean 
$\langle\xi(r)\rangle$=0 and dispersion $\xi$. 
For illustration in Fig.~\ref{rho} we show the 
numerical profile generated using this procedure 
for $L_0= 700$~km and $\xi= 0.1$. 

Substituting this realization of the matter density into the Schr\"{o}dinger 
equation for two neutrino flavors we have solved 
the Cauchy problem for different starting points $r_0$ within 
the solar core $r_0\leq R_{core}=0.3R_{\odot}$ with the same initial 
condition  $\nu_e(r_0)=1$. 
The production points $r_0$ are chosen to be placed in knots of a  
$30\times 60$-net that covers the cross-section of the core hemisphere, 
$r< 0.3R_{\odot}$. In other words, $0.01R_{\odot}\simeq 7000$ km is the 
cell size of a k-rectangle chosen within core. 

In this way, for $30\times 60=1800$ $r_0$-points, we have obtained a set 
of the complex wave functions,  $\nu_a(r,r_0)= \mid \nu_a(r,r_0)\mid 
\exp (i\Phi (r,r_0) )$, a=e,x, from which one can easily 
get the survival probability at the surface of the Sun, $P_{ee}(R_{\odot},r_0)
\equiv P_{ee}(R_{\odot},r_0, s^2_2,\delta,L_0,\xi ) $ and 
the survival probability on the 
day--side of the Earth after propagation in vacuum through the solar wind
(e.g. for $\nu_{e}\to\nu_{\mu }$ oscillations),
\begin{eqnarray}
P_{ee}^{day}(r_0)&\equiv & P_{ee}^{day}(r_0,s^2_2, \delta,L_0,\xi ) 
= P_{ee}(R_{\odot},r_0) +\frac{s_2^2}{2}[1 - 2 P_{ee}
(R_{\odot},r_0)] -\nonumber\\ 
&&- \frac{1}{2}s_2c_2[\nu_{e}^*(R_{\odot},r_0)\nu_{\mu }(R_{\odot},r_0) 
+ \nu_{\mu }^*(R_{\odot},r_0)\nu_{e}(R_{\odot},r_0) ]~.
\label{dayside}
\end{eqnarray}

Then, assuming spherical symmetry, we have averaged the survival probability 
in Eq.~(\ref{dayside}) over the production points $r_0$.
The averaging means the multiplication  by the weight factor defined as the 
local $\nu$-source distribution $S_i(r_{0k})$, given by SSM BP98 for each  
$r_{ok}$ and each neutrino flux type i=pp, Be, pep,..., resulting in:
\begin{equation}
\langle P^{day,i}_{ee}(s^2_2,~\delta,L_0,\xi)\rangle 
= \frac{1}{1800}\sum_{k=1}^{1800} P_{ee}^{day}(r_{0k})S_i(r_{0k})~.
\label{average}
\end{equation}
Notice that Eq.~(\ref{dayside}) is equivalent to the standard expression
for the day--side survival probability 
$P_{ee}^{day}(r_0) = P_{e1}^{Sun}(r_0)P_{1e}^{Earth}+  P_{e2}^{Sun}(r_0)
P_{2e}^{Earth}= 
\mid c^2\psi_e - cs\psi_{\mu}\mid^2 + \mid s^2\psi_e + cs\psi_{\mu}\mid^2$ 
where the complex wave functions $\psi_a = \nu_a(R_{\odot}, r_0 )$  are 
given at the surface of the Sun and during the day 
$P_{1e}^{Earth}=1-P_{2e}^{Earth}=\cos^2\theta$.
During the night, solar neutrinos cross the Earth before reaching the detector 
and regeneration of $\nu_e$'s is possible \cite{daynight}. 
In order to take into account this effect we compute the
probability $P_{2e}^{Earth}$ by integrating numerically the differential 
equation that describes the evolution of neutrino flavors in the Earth.
This probability depends on the amount of Earth matter travelled by
the neutrino, or in other words, in its arrival direction which is
usually parametrized in terms of the zenith angle $\Phi$.
Thus, in general, the survival probability for a neutrino of given  
source $i$ arriving at a given zenith angle $\Phi$ is given by  
\begin{eqnarray}
\langle P_{ee}^{\Phi,i}(s_2^2,\delta,L_0,\xi) 
\rangle &=& \langle P_{ee}^{day,i}(s_2^2,\delta,L_0,\xi) \rangle 
\nonumber  \\ 
&  & + \frac{(2\langle P_{ee}^{day,i}
(s_2^2,\delta,L_0,\xi) \rangle - 1)
\left (s^2 - P^{Earth}_{2e}(s_2^2,\delta,L_0,\xi;\Phi) \right)}{cos 2\theta}~,
\label{finalprob}
\end{eqnarray}
Such probability remains a function of the 
fundamental neutrino parameters $\delta$ and $s_2^2$,
as well as of two noise parameters $\xi$, $L_0$. 

In principle we should also average over different realizations of the
density perturbations with the same level of noise $\xi$ and correlation 
length $L_0$.  We discuss next that our averaging procedure over
the grid of production points $r_0$  is equivalent to the numerical average 
over an ensemble of electron densities. 

Notice that for parallel rays which are directed along the z-axis to the Earth 
at a fixed distance $z=z_0$ from the center, 
the density profile Eq.~(\ref{noisedensity}) 
has different density amplitudes since
$z=r_1$ only for one ray in equiliptics, for other
k rays ($k\neq 1$) the hypotenuse is longer, $r_k >z$. Thus, rays are not 
equivalent to each other.
This means that considering parallel rays and for any k-ray 
substituting the same 
final distance $z=R_{\odot}$ into the wave functions 
$\nu_a(R_{\odot}, r_{0k})$ we automatically took into account different 
density profile realizations including a matter noise.
Then integrating (summing) 
over $r_0$ in Eq.~(\ref{average}) we have averaged simultaneously over 
different realizations of noise.

Alternatively we may think of our averaging procedure in the following
way. In our distribution of starting points in the grid 
we have several points $j$ with the same value of $r_0$ but located 
at different distances $y_j$ from the Sun--Earth $z$ axis. All these points
encounter different realizations of the matter density in their way
as they move in their path parallel to the $z$ axis since 
for each of them $r_j(t)=\sqrt{z_j(t)^2+y_j^2}$ is different and so it is
the profile $\rho$ they are subject to at each time $t$. 
So our averaging over 1800 initial points can be understood as an 
integration over the starting point $r_0$ times an average over different 
matter density realizations $j$ for each $r_0$.

In Fig.\ref{probdm} we show the averaged survival probabilities
$<P^{day}_{ee}(s_2^2,\delta,L_0,\xi)>$ 
averaged for the $^8B$ production point distribution,  
as a function of $\delta=\Delta m^2/4E$ for two values of the mixing
angle $s_2^2 = 0.006$ (SMA) and $s_2^2 =.78$ (LMA) for level noise $
\xi=4$ \% and different values of the correlation length $L_0$.
Also shown in the figure is the corresponding probability for 
the noiseless case.
As shown in the figure, even for modest noise level $\xi=4$ \% 
we find relatively large effects. This is specially the case for
the SMA value. For LMA the effect is large only for short
correlation lengths. We next discuss the interpretation of
this behaviour. 

\subsection{Parametric resonance for MSW conversions in noisy matter}
\label{resonance}

In Fig.~\ref{contours} we show isolines 
$\Delta P_{ee}=
[<P^{day}_{ee}(s^2_2,\delta,L_0,\xi)> - <P^{day}_{ee}(s^2_2,\delta)>
]$ (averaged for the $^8B$ production point distribution)  
in the plane  $\delta$, $L_0$ for noise level $\xi=4$ \%. 
$P^{day}_{ee}(s^2_2,\delta)$ is the survival probability in the
absence of noise. Figure \ref{contours}.a corresponds to fixed 
$s_2^2=0.0063$ while Fig.~\ref{contours}.b corresponds to $s_2^2 = 0.79$. 

One can see in Fig.~\ref{contours}.a a wide spectrum of the domain sizes 
$L_0$, for which the noisy solution  gives large difference for the 
$\nu_e$- suppression, $\Delta P_{ee}= 0.3-0.5$, even 
for a modest noise level (=4 \% )!. Largest enhancement occurs for
values of $\delta=(1$--3$)\times 10^{-6}$ eV$^2$/MeV. 
For characteristic values of  
$\Delta m^2\sim (0.4$--$1)\times10^{-5}~eV^2$ in the SMA region the 
energy values for such enhancement in Fig.~\ref{contours}.a
corresponds to $E = \Delta m^2(eV^2)/(4\delta)$~MeV $\simeq
0.33$--2 MeV in the interesting range for the existing solar neutrino
experiments.

For the larger mixing angle the enhancement is smaller 
(see Fig.~\ref{contours}.b). The maximum  effect of about 
$(\sim 0.13)$ appears somewhere near $L_0\sim 70-100$ km.

For both cases we can interpret this enhancement as being due to 
{\it parametric effects in 
neutrino oscillations}. The parametric resonance implies a synchronization 
between the system eigen--oscillations having MSW frequencies 
$\omega = 2\pi/l_m$
and the parameter variations given by a changing size of density 
fluctuations $L_0$, $\omega = 2\pi/L_0$. Here the neutrino oscillation 
length in medium $l_m$ is given by
\begin{equation}
l_m (r)= \frac{l_{\nu}}{\sqrt{(\cos 2\theta - l_{\nu}/l_0(r))^2 + 
\sin^22\theta}}~,
\label{medium}
\end{equation}
where $l_{\nu}= 4\pi E/\Delta m^2$ is the oscillation length in vacuum, 
$l_0(r) = 2\pi m_p/\sqrt{2}G_F\rho (r)$ is the refraction length.

The synchronization condition (parametric resonance condition) states that the
phase acquired by the oscillating neutrino state on one fluctuation length 
$L_0$ should be a multiple of 2$\pi$ \cite{Krastev}
\begin{equation}
\int_0^{L_0}dr\frac{2\pi}{l_m(r)}= 2\pi k,~k=1,2,3,...
\label{param}
\end{equation}
One can simplify this condition  
for SMA MSW when the neutrino energy differs considerably from 
the MSW resonance energy, 
$\mid \cos 2\theta - l_{\nu}/l_0\mid \gg \sin 2\theta$.  
In this approximation substituting the mean density $\rho = \bar{\rho}$ 
we obtain from Eq.~(\ref{param}) the simple formula for parametric resonance 
in the case of SMA MSW  oscillations \cite{Chechin}
\begin{equation}
\mid \frac{\cos 2\theta}{l_{\nu}} - \frac{1}{\bar{l_0}}\mid = \frac{k}{L_0}~.
\label{param2}
\end{equation}

According to Eq.~(\ref{param2}), two types of resonance regimes may appear 
depending on the value of $L_0$ :
\begin{itemize}
\item A resonance at low energy when $\delta\gg \delta_R = 
G_F\rho_R/(\sqrt{2}m_p\cos 2\theta )$ ($\bar{\rho}\leq \rho_R$).
In this case, for $L_0\ll \bar{l_0}$,  
one obtains the parametric resonance condition 
$l_m\simeq l_{\nu}\simeq L_0$.
\item A resonance at energy well above the MSW resonance, 
$\delta \ll \delta_R$ 
($\bar{\rho}> \rho_R$). In this case the parametric resonance condition is 
$l_m\simeq l_0\simeq L_0$.
\end{itemize}

A more interesting case occurs when the synchronization condition 
in Eq.~(\ref{param}) is achieved in the proximity of the MSW resonance
($\delta = \delta_R$). In this case we can substitute 
$l_m^{res} = l_{\nu}/\sin 2\theta$, and still assuming a short correlation 
length, $L_0< \Delta r$, compared to the thickness of the resonant layer 
$\Delta r$, we get much longer $L_0$-sizes 
\begin{equation}
L_0 = kl_m^{res} = \frac{k\pi}{\delta_R\sin2\theta}.
\label{param3}
\end{equation} 
The parametric resonance is stronger for this case \cite{Krastev}. 

Since for SMA MSW conversions the thickness of the resonant layer is
$\Delta r = \rho(d\rho/dr)^{-1}/\cot 2\theta \simeq 0.1R_{\odot}/
\cot 2\theta\sim 7000$ km the condition in Eq.~(\ref{param3}) remains valid
for some first numbers k = 1,2,\dots . In other words, the condition  
7000~km $\gtrsim L_0 = k\pi/\delta_R \sin 2\theta$ implies a lower bound
on the possible values of $\delta_R$ for which the parametric resonance
is possible. For instance for the mixing angle in Fig.~\ref{contours}.a
$\sin(2\theta)=0.08$, the value $\delta_R\simeq 10^{-6}$ eV$^2$/MeV is
only marginally allowed for $k=1$.

Conversely from the MSW condition 
$l_m^{res} = 250$ km(E/MeV)$/(\Delta m^2/10^{-5}$ eV$^2$)$/\sin 2\theta )$ 
substituting the SMA MSW parameter values $\Delta m^2 \sim 10^{-5}$ eV$^2$, 
$\sin 2\theta\sim 0.1$ we find $l_m^{res,SMA}\simeq 2500 (E$/MeV)~km.
Moreover we see in Fig.~\ref{contours}.a 
that domain sizes 
$L_0= 700-2000$~km are appropriate for maximum enhancement which implies 
that the MSW resonant parametric condition Eq.~(\ref{param3}) is mainly 
achieved for $k =1$ 
in the case of {\it low neutrino energies} like the pp-neutrinos 
seen in GALLEX and SAGE. 
 
Let us note that the parametric resonant condition in Eq.~(\ref{param3}) is 
rather opposite to what was assumed in Eq.~(3.18) \cite{Nunokawa}, 
$L_0 = 0.1\;l_m$. In other words, $\delta$-correlated noise,
$L_0\ll l_m$, is an analytic approximation which is appropriate only for 
small values of the correlation length. To illustrate this we
plot in Fig.~\ref{probl} 
the survival probability 
$\langle P_{ee} (s^2_2,\delta,L_0,\xi)\rangle$ as the function of $L_0$ 
for different values of the level noise and fixed 
$\delta=10^{-6}$ eV$^2$/MeV. Figure~\ref{probl}.a corresponds to  
$s_2^2= 0.01$ ($l_m^{res}\simeq 7000$ km) 
and Fig.~\ref{probl}.b to $s_2^2 = 0.7$ ($l_m^{res}\simeq 700$ km).  

One can see in Fig.~\ref{probl} different heights of the ``bumps'' of 
the survival probability for different noise levels whereas the position
of the peaks are similar cutting sharply at the right edge 
$L_0 = l_m^{res}$ - 
position of the strongest parametric resonance Eq.~(\ref{param3}) for $k=1$. 
If $\delta \neq \delta_R$, or energy E is far from the MSW resonance 
value $E_R$
the parametric resonance still takes place but from corresponding 
Eq.(\ref{param2}) we find that lower $L_0 \sim l_m < l_m^{res}$ are 
appropriate. 
Only if $L_0\ll l_m$ the $\delta$-correlated regime starts.

It is interesting to compare our results in Fig.~\ref{probl}.a 
with the numerical results obtained for the ``Cell'' model plotted in 
Fig.~1 of Ref.~\cite{Burgess} where authors averaged the ordinary
Parke formula over random ensemble of 200 density profiles of Cell type
(dashed, dot-dashed and solid thick lines on that figure).  
These 200 profiles correspond to 200 neutrino creation sites within 
the core only in contrast with 1800 points in our {\it direct 
numerical simulation of the Schr\"{o}dinger equation} Eq. (\ref{master}).

Notice that in that figure 
$\delta=2 \delta_{Fig.~\ref{probl}}$.
We find that both figures present similar behaviour in the short
correlation length regime and a comparable enhancement for 
$L_0\sim 10^2$--$10^4$ which ends at $L_0\simeq l^{res}_m$.
But for $L_0\gg l^{res}_m$ the results in 
Fig.~1 of Ref.~\cite{Burgess} present a strong dependence on the 
position of the resonance within the cell.
The result of our numerical calculation is closer to the 
``Cell'' model result with randomly distributed cell positions 
(dot-dashed line in Fig.~1 of Ref. \cite{Burgess}) and the mean 
$\langle P_{ee}(L_0)\rangle$ tends to the noiseless MSW survival 
probability as expected for large $L_0$.  

Concluding this section we want to remark that the numerical approach 
presented here with the averaging 
of the solutions $P_{ee}(r_0)$ over noise realizations is quite different 
from any previous approaches with the averaging
of the Schr\"{o}dinger equation itself before obtaining of a solution.
Only the straightforward numerical solution of the Schr\"{o}dinger equation 
is an appropriate way to tackle the problem in the general case of
arbitrary correlation lengths.
\section{Fits: Results}
\label{analysis}
\subsection{Data and Techniques} 
\label{data}
In order to study the possible values of neutrino masses and mixing 
for the oscillation solution of the solar neutrino problem 
in the presence of noisy matter density fluctuations in the sun, we have 
used data on the total event rates measured in the Chlorine experiment at 
Homestake \cite{homestake}, in the two Gallium experiments GALLEX and 
SAGE \cite{gallex,sage} and in the water Cerenkov detectors Kamiokande and 
Super--Kamiokande shown in Table~\ref{rates}. Apart from the total event 
rates, we have in this last case 
the zenith angle distribution of the events and the electron recoil 
energy spectrum, all measured 
with their recent 825-day data sample~\cite{sk99}. 
 
For the  calculation of the theoretical expectations we use the BP98 standard  
solar model of Ref.~\cite{BP98}.  
The general expression of the expected event rate in the presence of 
oscillations in experiment $i$ is given by $R^{th}_i$ : 
\begin{eqnarray} 
R^{th}_i & = & \sum_{k=1,8} \phi_k 
\int\! dE_\nu\, \lambda_k (E_\nu) \times  
\Big[ \sigma_{e,i}(E_\nu)  \langle  
P^k_{ee}(s^2_2,\delta,L_0,\xi)
\rangle \label{ratesth} \\ 
& &                            + \sigma_{x,i}(E_\nu)  
\bigg(1-\langle P^k_{ee}(s^2_2,\delta,L_0,\xi)
\rangle \bigg)\Big] \nonumber. 
\end{eqnarray}   
where $E_\nu$ is the neutrino energy, $\phi_k$ is the total neutrino 
flux and $\lambda_k$ is the neutrino energy spectrum (normalised to 1) 
from the solar nuclear reaction $k$ with the 
normalization given in Ref.~\cite{BP98}. Here $\sigma_{e,i}$ 
($\sigma_{x,i}$) is the $\nu_e$ ($\nu_x$, $x=\mu,\,\tau$) interaction 
cross section in the Standard Model with the target 
corresponding to experiment $i$. 
For the Chlorine and Gallium experiments we use improved cross 
sections $\sigma_{e,i}(E)$ from 
Ref.~\cite{prod}. For the Kamiokande and Super--Kamiokande experiment  
we calculate the expected signal with the corrected cross section as  
explained below. 
$\langle P_{ee}^k \rangle$ is  
is the yearly averaged $\nu_e$ survival probability at the detector, 
as given by Eq.~(\ref{finalprob}) after averaging over arrival directions
$\Phi$.
 Note that  $\langle P_{ee}^k \rangle$ is a function of the oscillation 
parameters as well as the noise
parameters. 

We have also included in the fit the experimental results from the 
Super--Kamiokande Collaboration on the zenith angle distribution of 
events taken on 5 night periods and the day averaged value 
~\cite{sk99}. We compute the expected event rate  
in the period $a$ in the presence of MSW oscillations as, 
\begin{eqnarray}  
R^{th}_{sk,a}  
& = & \frac{\displaystyle 1}{\displaystyle\Delta \tau_a} 
\int_{\tau(\cos\Phi_{min,a})}^{\tau(\cos\Phi_{max,a})}  d\tau 
\sum_{k=1,8} \phi_k\int\! dE_\nu\, \lambda_k (E_\nu) \times  
\Big[ \sigma_{e,sk}(E_\nu)  
\langle P^k_{ee}(s^2_2,\delta,L_0,\xi;\tau) \rangle   
\label{eq:daynight}\\  
& &+ \sigma_{x,sk}(E_\nu)  
\bigg(1-\langle P^k_{ee}(s^2_2,\delta,L_0,\xi;\tau) \rangle   
\bigg)\Big] \nonumber 
\,, 
\end{eqnarray}   
where $\tau$ measures the yearly averaged length of the period $a$  
normalized to 1, so $\Delta\tau_a=\tau(\cos\Phi_{max,a})-\tau 
(\cos\Phi_{min,a})=$ 
.500, .086, .091, .113, .111, .099 for the day and five night periods. 
Notice that the dependence of 
$\langle P^i_{ee}(s^2_2,\delta,L_0,\xi;\tau) \rangle$ on $\tau$ comes
only from the dependence of the Earth regeneration probability 
$P^{Earth}_{2e}$ on the different Earth matter profile crossed by 
the neutrino during the five night periods.
 
The Super-Kamiokande Collaboration has also measured the recoil 
electron energy spectrum.  In their published analysis \cite{sk1} 
after 504 days of operation they present their results for energies 
above 6.5 MeV using the Low Energy (LE) analysis in which the recoil 
energy spectrum is divided into 16 bins, 15 bins of 0.5 MeV energy 
width and the last bin containing all events with energy in the range 
14 MeV to 20 MeV.  Below 6.5 MeV the background of the LE analysis 
increases very fast as the energy decreases. Super--Kamiokande has 
designed a new Super Low Energy (SLE) analysis in order to reject this 
background more efficiently so as to be able to lower their threshold 
down to 5.5 MeV. In their 825-day data \cite{sk99} they have used the 
SLE method and they present results for two additional bins with 
energies between 5.5 MeV and 6.5 MeV. 
In our study we use the experimental results from the 
Super--Kamiokande Collaboration on the recoil electron spectrum divided in 
18 energy bins, including the results from the LE analysis for the 16 
bins above 6.5 MeV and the results from the SLE analysis for the two 
low energy bins below 6.5 MeV.   
The general expression of the expected rate in a bin in the presence of 
oscillations, $R^{th}$, is similar to that in Eq.~(\ref{ratesth}), 
with the substitution of the cross sections with the corresponding 
differential cross sections 
folded with the finite energy resolution function of the detector 
and integrated over the electron recoil energy interval of the bin, 
$T_{\text {min}}\leq T\leq T_{\text {max}}$: 
\begin{equation} 
\sigma_{\alpha,sk}(E_\nu)=\int_{T_{\text {min}}}^{T_{\text {max}}}\!dT 
\int_0^{\frac{E_\nu}{1+m_e/2E_\nu}} 
\!dT'\,Res(T,\,T')\,\frac{d\sigma_{\alpha,sk}(E_\nu,\,T')}{dT'}\ . 
\label{sigma} 
\end{equation} 
The resolution function $Res(T,\,T')$ is of the form~\cite{sk1,Faid}: 
\begin{equation} 
Res(T,\,T') = \frac{1}{\sqrt{2\pi}(0.47 \sqrt{T'\text{(MeV)}})}\exp 
\left[-\frac{(T-T')^2}{0.44\,T' ({\text {MeV}})}\right]\ , 
\end{equation} 
and we take the differential cross section $d\sigma_\alpha(E_\nu,\,T')/dT'$  
from \cite{CrSe}. 
 
In the statistical treatment of all these data we perform a $\chi^2$ 
analysis for the different sets of data,  
following closely the analysis of Ref.~\cite{fogli-lisi} with the  
updated uncertainties given in Refs.~\cite{lisi3,BP98,prod}, as 
discussed in Ref.~\cite{oursolar}.  We thus define  
a $\chi^2$ function for the three set of observables  
$\chi^2_{\text {rates}}$, $\chi^2_{\text {zenith}}$, and  
$\chi^2_{\text {spectrum}}$ where in both $\chi^2_{\text {zenith}}$, and 
 $\chi^2_{\text {spectrum}}$ we allow for a free normalization in order to  
avoid double-counting with the data on the total event rate which is 
already included in $\chi^2_{\text {rates}}$.   
In the combinations of observables we define the 
$\chi^2$ of the combination as the sum of the different $\chi^2$'s.  
In principle such  analysis should be taken with a grain of salt as 
these pieces of information are not fully independent; in fact, they 
are just different projections of the double differential spectrum of 
events as a function of time and energy. Thus, in our combination we 
are neglecting possible correlations between the uncertainties in the 
energy and time dependence of the event rates. 

\subsection{MSW Regions}
We present next the results of the allowed regions in the two--parameter 
space $\Delta m^2$, $\sin^2(2\theta)$ for the analysis of 
the different combination of observables. 
In building these regions, for a  
given set of observables and a certain value of the noise parameters
$L_0$ and $\xi$ (in what follows we present the results for 
$\xi=4\%$ which is a reasonable large value for the noise amplitude)
we compute for any point in the parameter space  
of two--neutrino oscillations 
the expected values of the observables and with those and the corresponding 
uncertainties we construct the function  
$\chi^2(\Delta m^2,\sin^2(2\theta);L_0,\xi)_{obs}$.  
We find its minimum in the full two-dimensional space of 
MSW oscillations. The allowed regions  
for a given CL are then defined as the set of points satisfying  
the condition:  
\begin{equation} 
\chi^2(\Delta m^2,\sin^2(2\theta);L_0,\xi)_{obs}  
-\chi^2_{min,obs}(L_0,\xi)\leq \Delta\chi^2 \mbox{(CL, 2~dof)} 
\label{deltachi2} 
\end{equation}  
where, for instance $\Delta\chi^2($CL, 2~dof)=4.6, 6.1, and 9.2 for 
CL=90, 95, and 99 \% respectively. In  
Figs.~\ref{rates}--\ref{global} we plot the corresponding allowed
regions for $\xi=4$ \% and for five values of the correlation length
$L_0=$70,~200,~700,~2000, and 10$^4$ km. For comparison in the last
panel we also show the regions in the absence of random noise.

Figure~\ref{rates} shows the results of  the fit to the observed total 
rates only. We see in the figure that for any value of the correlation
length we always find the three allowed regions, SMA, LMA and LOW
as in the standard noiseless MSW analysis although their extend
in $\Delta m^2$ and $\sin^2(2\theta)$ varies with the value of the
correlation length $L_0$. The presence of noise modifies the 
shape and size of the allowed regions through two different
although related effects. Due to the modification of the shape of 
the electron survival probability, the value of the expected rates
for a given point on the MSW plane are different once the noise
is included what modifies the value of $\chi^2$ for that given 
point. This leaves also to a shift on the value
of $\chi^2_{min}$ used to define the regions.
 
In Table~\ref{chimin} we show the values of the local $\chi^2 _{min}$ 
in the three regions for the different values of the
correlation length. 
First thing we notice is that for the analysis of the rates only 
in the LMA region
the value of $\chi^2 _{min}$  is always lower in the presence of
noise. This can be understood by looking at Fig.~\ref{probdm}.b
where we see that noise increases the survival probability 
for $2\times 10^{-7}\lesssim \delta \lesssim 3\times 10^{-6}$,
ie for lower values of the neutrino energies relevant for the 
Gallium experiments. This increases the rate at the Gallium experiments 
which is underestimated in the standard LMA solution. For the same 
reason the LMA become larger and shifted to lower $\Delta m^2$ values 
and it extends to smaller angles. For larger $L_0$, noiseless LMA solution 
is obtained.

For the SMA solution we find that 
unless for very short or very long correlation lengths, for which 
$\chi^2 _{min}$ 
is almost the same as in the absence of noise,  
for 100 km $\lesssim L_0\lesssim$ few 1000 km, the presence of noise
leads to an increase on the value of $\chi^2 _{min}$, or in other words
to a worse description of the data in the SMA region. In 
Fig.~\ref{probdm}.a we see that the presence of noise leads to an 
increase on the survival probability for
$2\times 10^{-7}\lesssim \delta \lesssim 3\times 10^{-6}$, so there is
not enough suppression of the Berilium neutrinos in the relevant
$\Delta m^2$ values of the SMA solution. For this reason the rate
at Chlorine experiment is increased and the fit worsens as compared 
to the noiseless case. The worsening is maximal for
$L_0$ of few thousand km when the effect of noise is maximally resonantly 
enhanced as discussed in Sec.~\ref{resonance}. In this case the LMA gives 
a better description of the measured rates. Also the SMA region is 
shifted towards larger values of the mixing angle.
Finally we notice that the LOW solution is basically unmodified by the
presence of noise as it mainly corresponds to values of 
$\delta\lesssim 10^{-7}$ which are little affected by the presence of
perturbations as seen in Fig.~\ref{probdm}.b.

\subsection{Zenith angle Dependence and Super--Kamiokande Spectrum}

Figure~\ref{ratesz} 
shows the regions allowed by the fit of 
both total rates and the Super--Kamiokande zenith angular distribution.
Also plotted is the excluded region at 99 \% CL from the 
zenith angular measurement. As seen in the figure the shape of the
excluded region is very little dependent of $L_0$, as expected. 
The zenith angular
dependence measures the regeneration effect on the neutrino survival
probability when the neutrinos travel through the Earth matter and it
is expected to be independent on the details of the sun matter modelling. 
The main effect of the inclusion of the day--night  
variation data is to cut down the lower part of the LMA region for
any value of $L_0$.
Since for short correlation lengths the LMA region had been shifted towards 
lower $\Delta m^2$ values, the inclusion of the zenith angle distribution
data leads to a reduction of the size of the LMA region for short $L_0$.
In this way, for instance, for $L_0=70$ km the LMA region at 99 \% CL extends 
only in the range $1.5\lesssim (\Delta m^2/10^{-5})$ eV$^2\lesssim 7$-- 
to be compared with $1.5\lesssim (\Delta m^2/10^{-5})$ eV$^2\lesssim 100$ 
in the absence of noise--. 
The SMA is also reduced in size as compared to the noiseless case.
We also find that once the zenith angle information is included 
the LMA becomes a better solution (lower $\chi^2_{min}$) for 
$L_0$=700--2000 km as seen in Table~\ref{chimin}. 
For these intermediate $L_0\sim 700-2000$ km, 
when the LMA region obtained from the analysis of the rates 
extended to smaller angles, some tiny regions 
in between the LMA and the SMA are still allowed 
at the 99 \% CL. 

In Fig.~\ref{ratesz} we show the regions allowed by the fit of 
both total rates and the Super--Kamiokande recoil electron 
energy spectrum.
Also plotted is the excluded region at 99 \% CL from the 
spectrum measurement. The main effect of the inclusion of the
spectrum data is to improve the quality of the LMA solution 
as compared to the SMA. Comparing with Fig.~\ref{rates} we
observe that the regions become larger after the inclusion of the
spectrum data. In particular the LMA region extends to larger
values of $\Delta m^2$. This behaviour is also observed in the
absence of noise. This is mainly due to the flattening of the
$\chi^2$ function after the inclusion of the spectrum data and
it is independent on the presence of noise. Notice that 
the larger sensitivity of the mean 
$\langle P_{ee}\rangle$ to the presence of noise occurs for low 
energy neutrinos as discussed in Sec.~\ref{solutions}. 
Having a threshold energy, $T_{th} = 5.5~MeV$, Super--Kamiokande 
is insensitive to such low energy effects.

\subsection{Global}

Figure~\ref{global} displays the results for the allowed regions
from the global analysis of the solar neutrino data including
the data on the total event rates, the zenith angular dependence
and the recoil electron energy spectrum. We find that, after 
all the observables are included the three regions remain valid.
The shape and size of final LMA solution  is very little affected
by the presence of noise. The SMA solution is maximally deformed
for correlation lengths few 100 km $\lesssim L_0\lesssim$ few 1000 km. 
It is for these values also that the SMA gives a worse description
of the observables. This behaviour is mainly driven by the 
effect on the total event rates which is the observable more
sensitive to the presence of noise. As discussed above both the
zenith angle distribution data and the spectrum data have very little
discriminating power on the noise parameters. One also finds that the
tiny 99\% CL ``islands'' in between the SMA and LMA are allowed
for  $L_0\simeq 700-2000$ km. 

Finally we notice that the LOW solution is basically unmodified by the
presence of the noise. For the neutrino energies accessible at 
existing experiments, the LOW region
corresponds to values of $\delta\lesssim 10^{-7}$ which are little affected 
by the presence of noise.

\section{Discussion}
\label{conclu}
In this paper we have studied the effect of density matter 
fluctuations in the sun on the MSW solutions to the SNP. 
Assuming no specific mechanism for generation of the fluctuations 
we have kept the amplitude and correlation length as independent parameters. 

Our analysis is performed under no assumption on the relative size
of the correlation length of fluctuations as compared to the 
neutrino oscillation length.
To perform such a study we have  solved numerically the evolution equation 
for the neutrino system including the full effect of the random matter density 
fluctuations of given amplitude and correlation length. 
Our procedure is to generate a realization of the density 
profile for given values of the perturbation amplitude and
correlation length and 
then to solve numerically the evolution equation for the
neutrino states for that given realization of the density profile and 
different neutrino production points and finally to average the obtained 
survival probability over different density realizations (with the same
amplitude and correlation length) and neutrino 
production points. This numerical approach 
of averaging  the solutions $P_{ee}(r_0)$ over noise realizations is 
different from any previous approaches with the averaging
of the Schr\"{o}dinger equation itself before obtaining the solution.

\begin{figure}
\begin{center}
\mbox{\epsfig{file=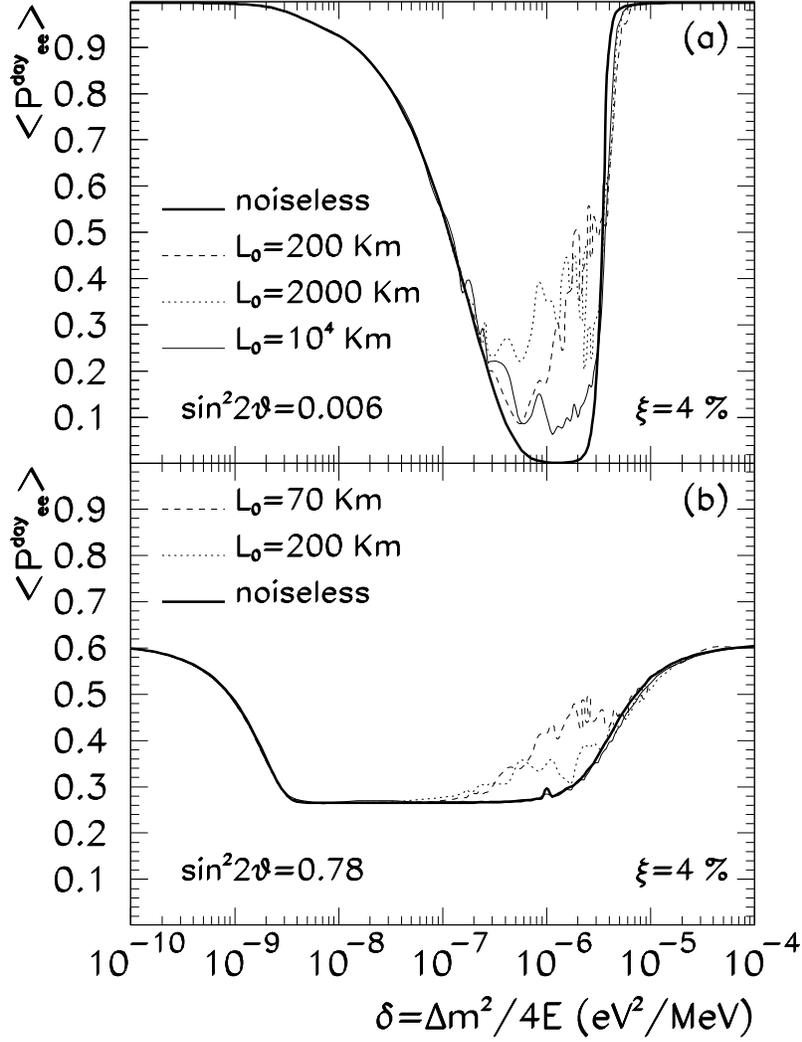,width=0.7\textwidth}}
\end{center}
\caption{Electron neutrino survival probability for $^8$B neutrinos
as a function of $\delta=\Delta m^2/(4E)$ for two values of the
mixing angle in the SMA region (a) $\sin^2(2\theta)=0.006$ and
in the LMA region (b) $\sin^2(2\theta)=0.78$ and for several 
values of the correlation length as label in the figure.
In both panels the level of noise $\xi=4$\%.}
\label{probdm} 
\end{figure}

\begin{figure}
\begin{center}
\mbox{\epsfig{file=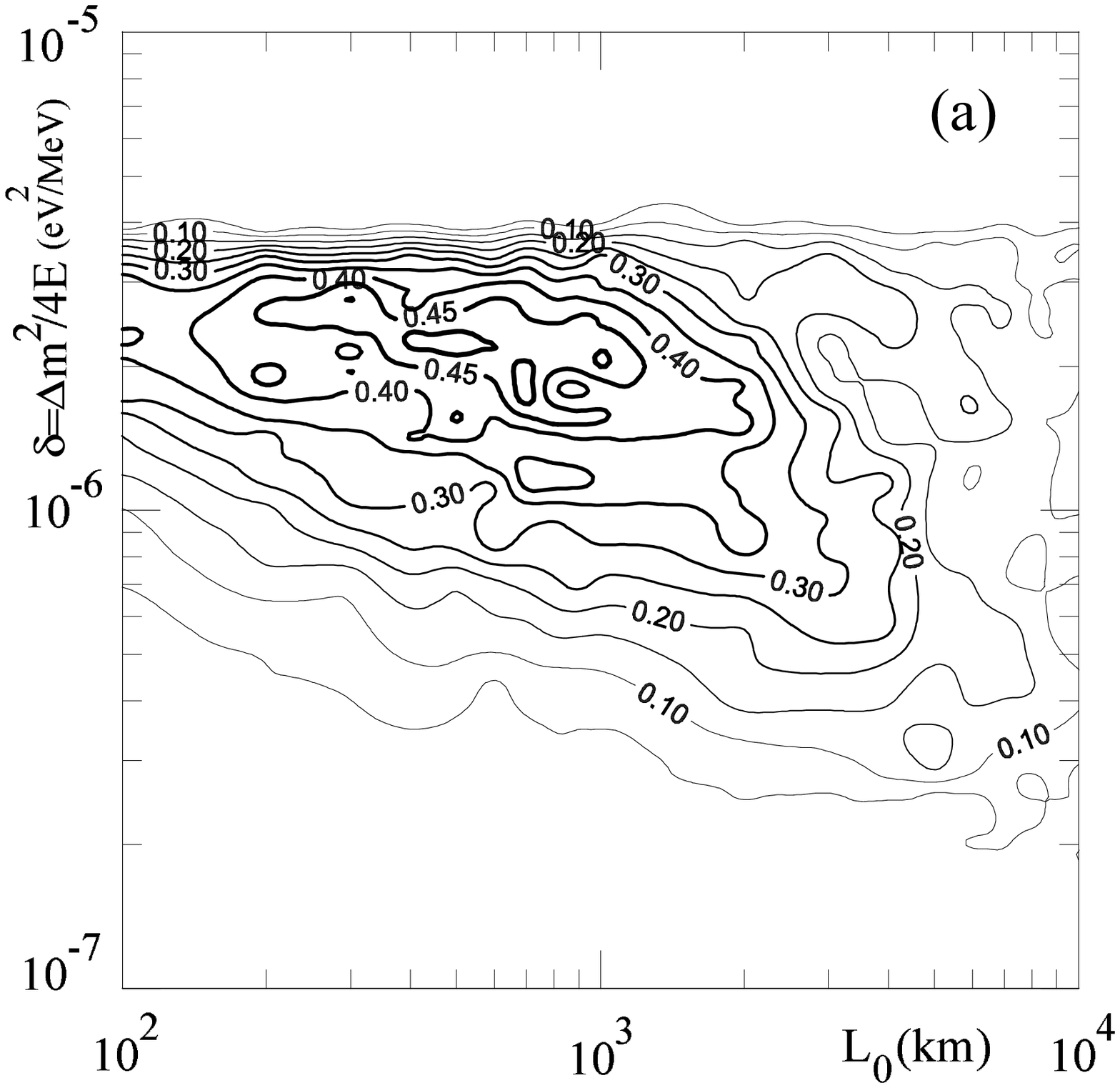,width=0.6\textwidth}}
\end{center}
\vspace{2.3cm}
\begin{center}
\mbox{\epsfig{file=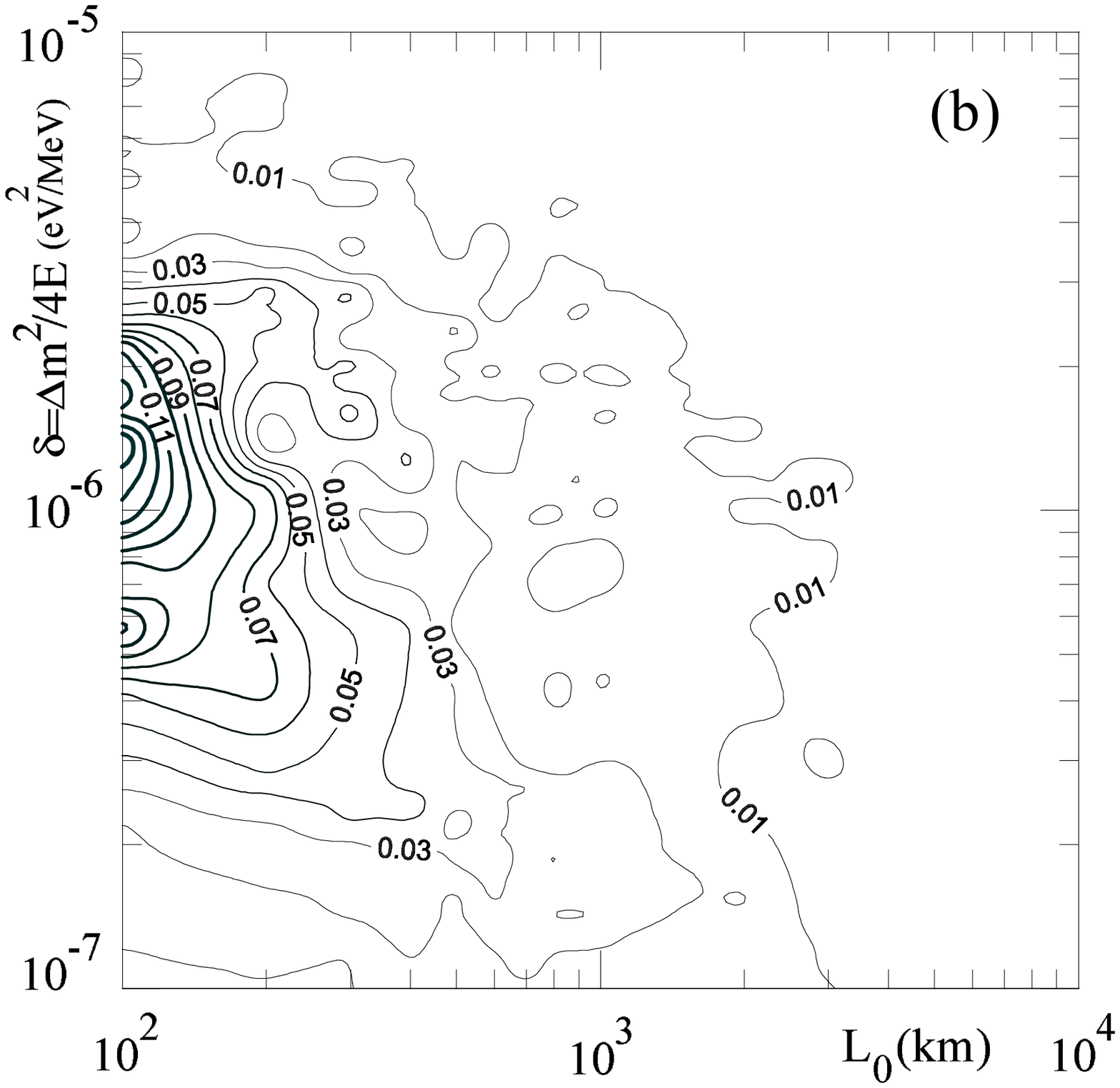,width=0.6\textwidth}}
\end{center}
\caption{Isocontours of constant           
$[<P_{ee}^{day}(L_0, \delta)> - P_{ee}^{day}(\delta)]$  values 
for $^8$B neutrinos in the 
plane $\delta$, $L_0$ for noise level $\xi$ = 4 \% and 
for two values of the mixing angle (a) $s_2^2=0.0063$ and 
(b)$s_2^2 = 0.79$. }
\label{contours} 
\end{figure}

\begin{figure}
\begin{center}
\mbox{\epsfig{file=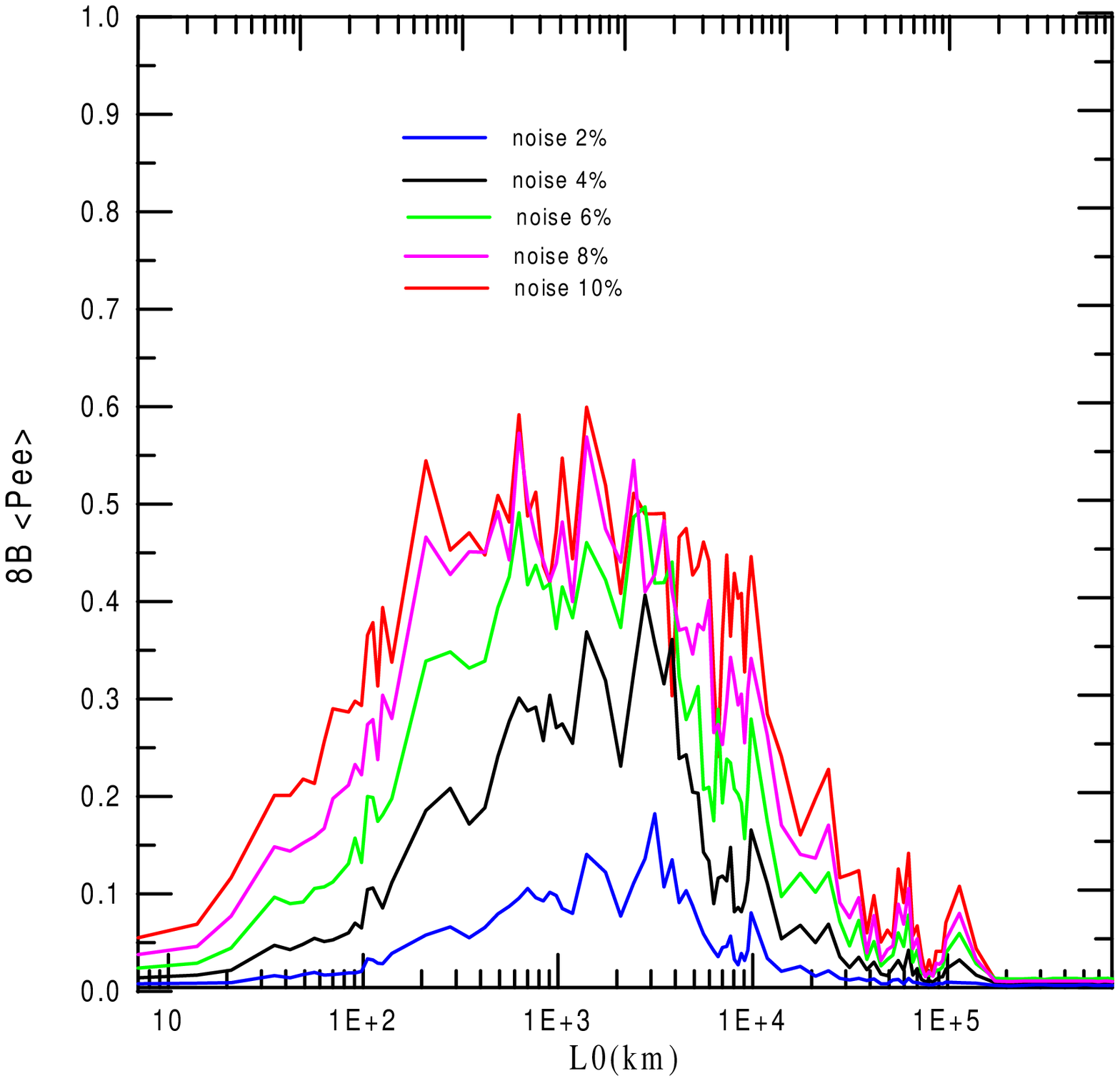,width=0.6\textwidth}}
\end{center}
\begin{center}
\mbox{\epsfig{file=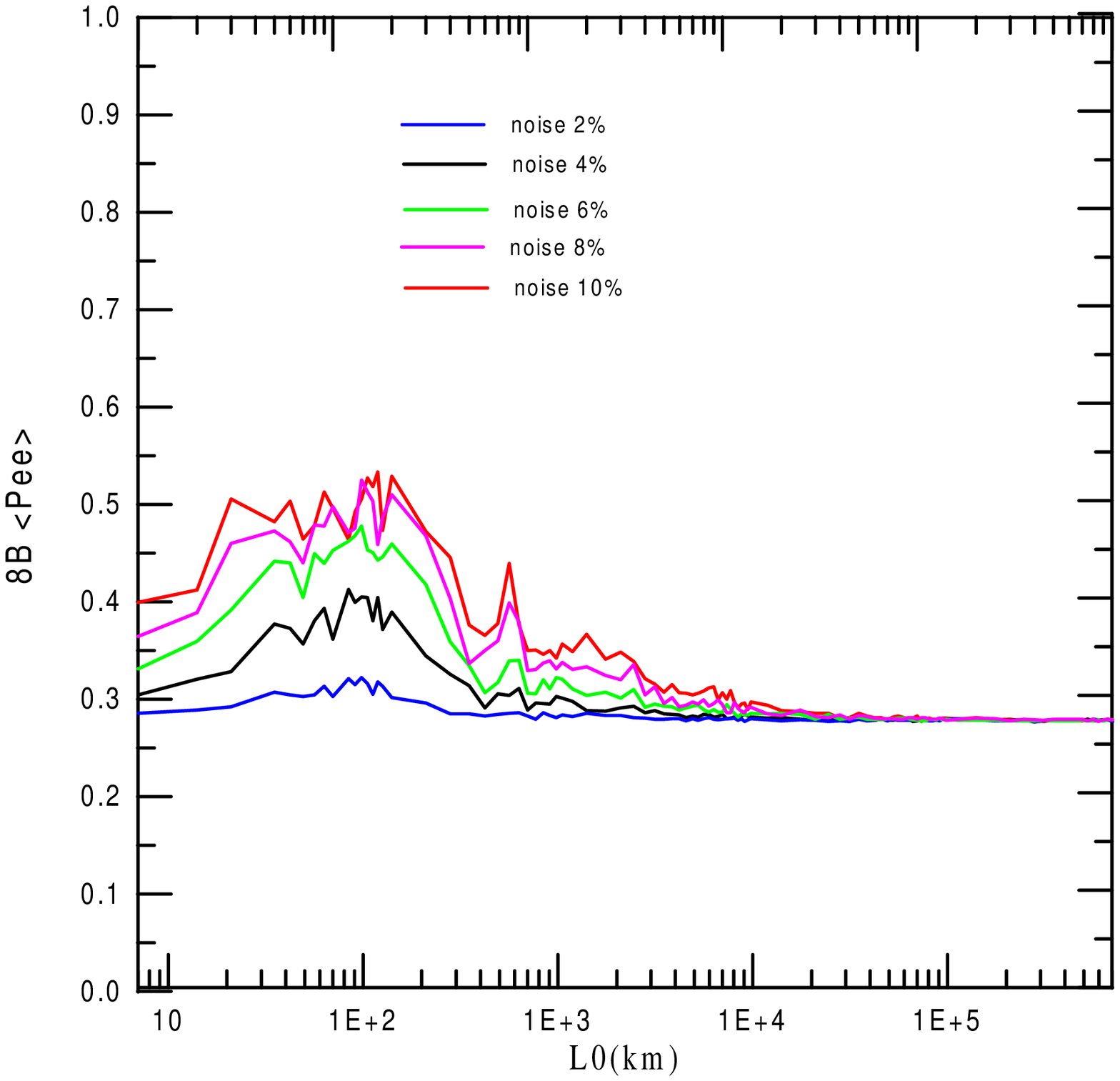,width=0.6\textwidth}}
\end{center}
\caption{Electron neutrino survival probability for $^8$B neutrinos
as a function of the correlation length $L_0$ for 
$\delta = 3.4 \times 10^{-6}$ for two values of the
mixing angle (a) $\sin^2(2\theta)=0.01$ and
(b) $\sin^2(2\theta)=0.7$ and for level of noise 
$\xi=$2 \%, \%, 8 \%, 10 \%.}
\label{probl} 
\end{figure}

Our results for the survival probabilities are presented in
Figs.~\ref{probdm}--\ref{probl}. We find that 
the effects are
larger for small mixing angles. The larger the mixing angle  
the shorter the correlation length needed
to observe an effect. For the SMA the larger effects occur for 
correlation lengths  in the range 
few 100 km $\lesssim L_0\lesssim$ few 1000 km. They can be understood
as due to a parametric resonance occuring when the
phase acquired by the oscillating neutrino state on one fluctuation length 
$L_0$ is a multiple of 2$\pi$. This resonance is maximal
when this condition is verified close to the MSW resonance.
We find that this resonant parametric condition is mainly 
achieved for {\it low neutrino energies} such as the pp-neutrinos 
seen in GALLEX and SAGE. 
 
\begin{figure}
\begin{center}
\mbox{\epsfig{file=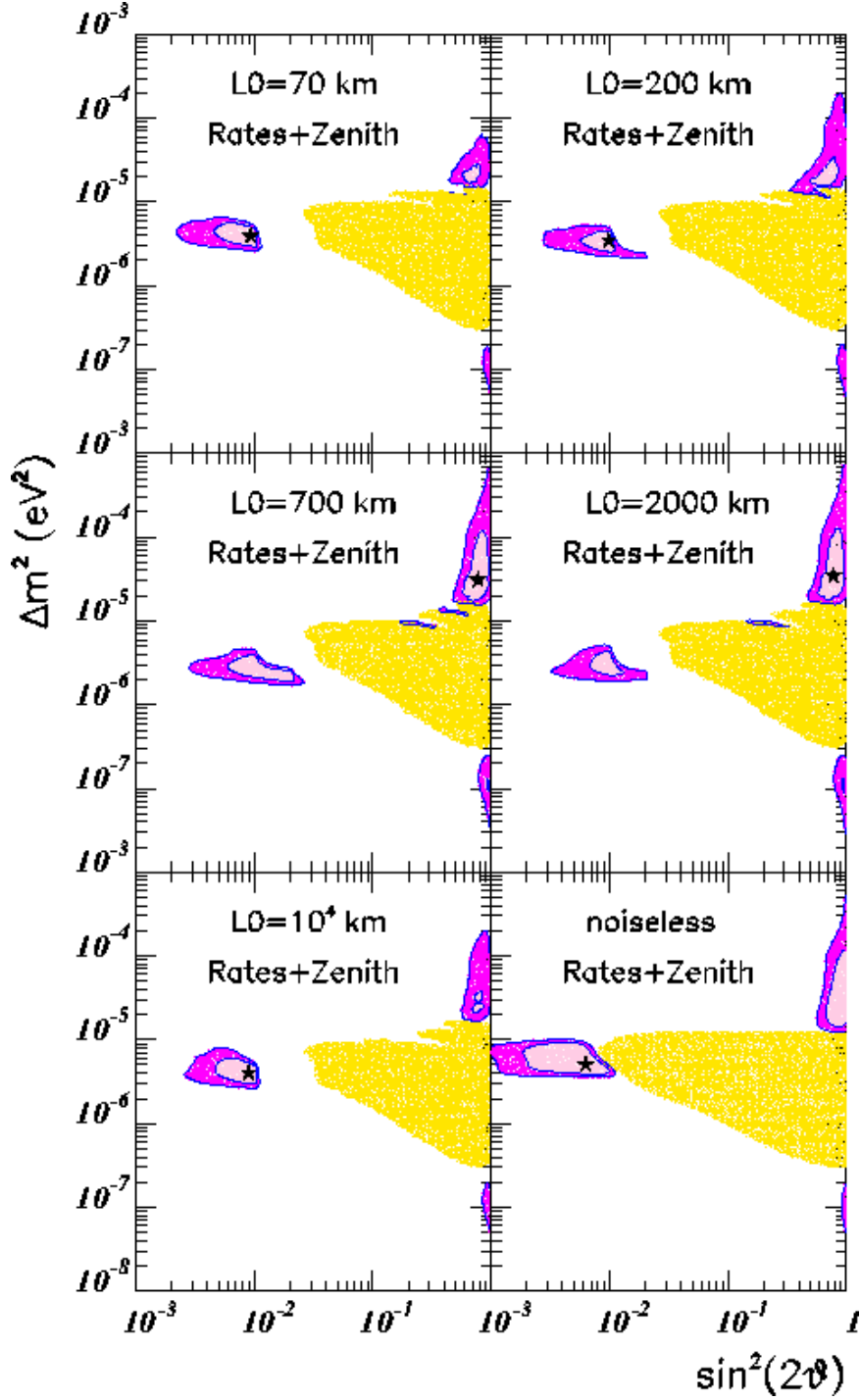,width=0.7\textwidth}}
\end{center}
\caption{Allowed regions in  $\Delta m^2$ and $\sin^2(2\theta)$ 
from the measurements of the total event rates at Chlorine, Gallium
and Super--Kamiokande (825-day data sample) for different correlation
lengths and for a noise level 
$\xi=$4 \%.  For the sake of comparison the
results for the standard noisless analysis is also shown.
he darker (lighter) areas indicate the 99\% (90\%)CL 
regions. The best--fit point used to defined the regions are 
indicated by a star.}
\label{rates}
\end{figure}

\begin{figure}
\begin{center}
\mbox{\epsfig{file=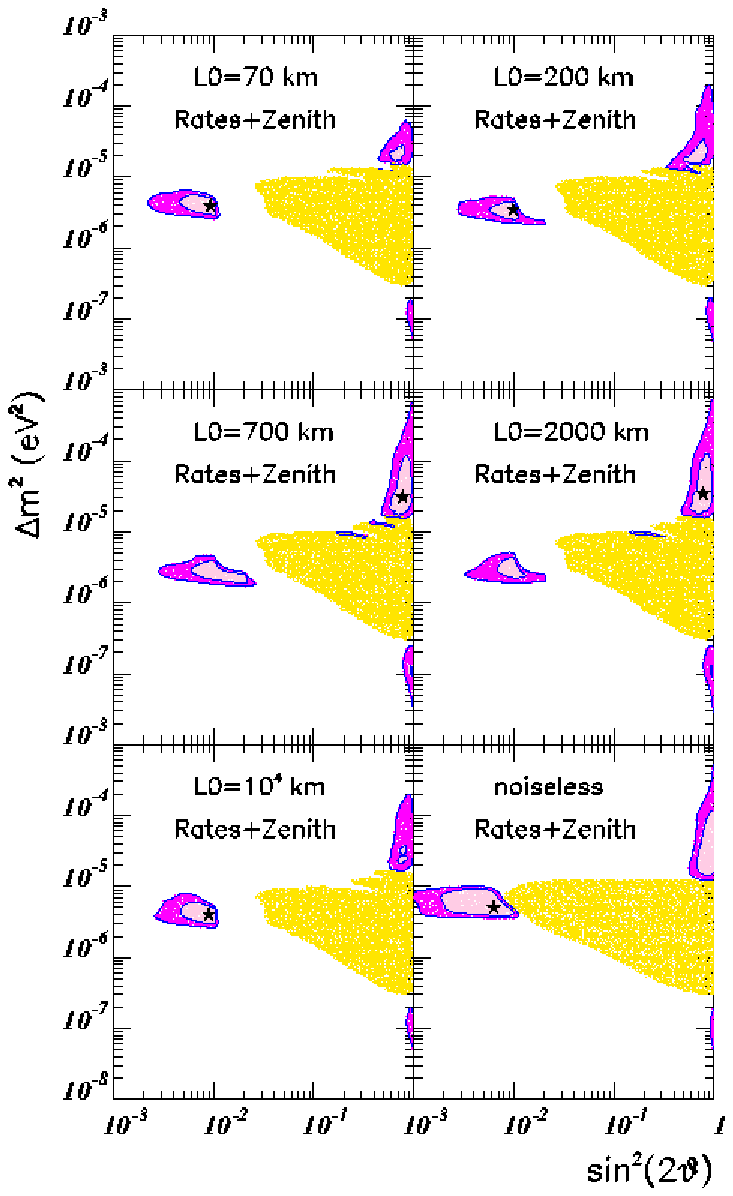,width=0.7\textwidth}}
\end{center}
\caption{Same as Fig.~\protect\ref{rates} 
but including also the data on the zenith
angle distribution observed in Super--Kamiokande. 
The shadowed area represents the
region excluded by the zenith angle distribution data at 99\% CL.} 
\label{ratesz} 
\end{figure}

\begin{figure}
\begin{center} 
\mbox{\epsfig{file=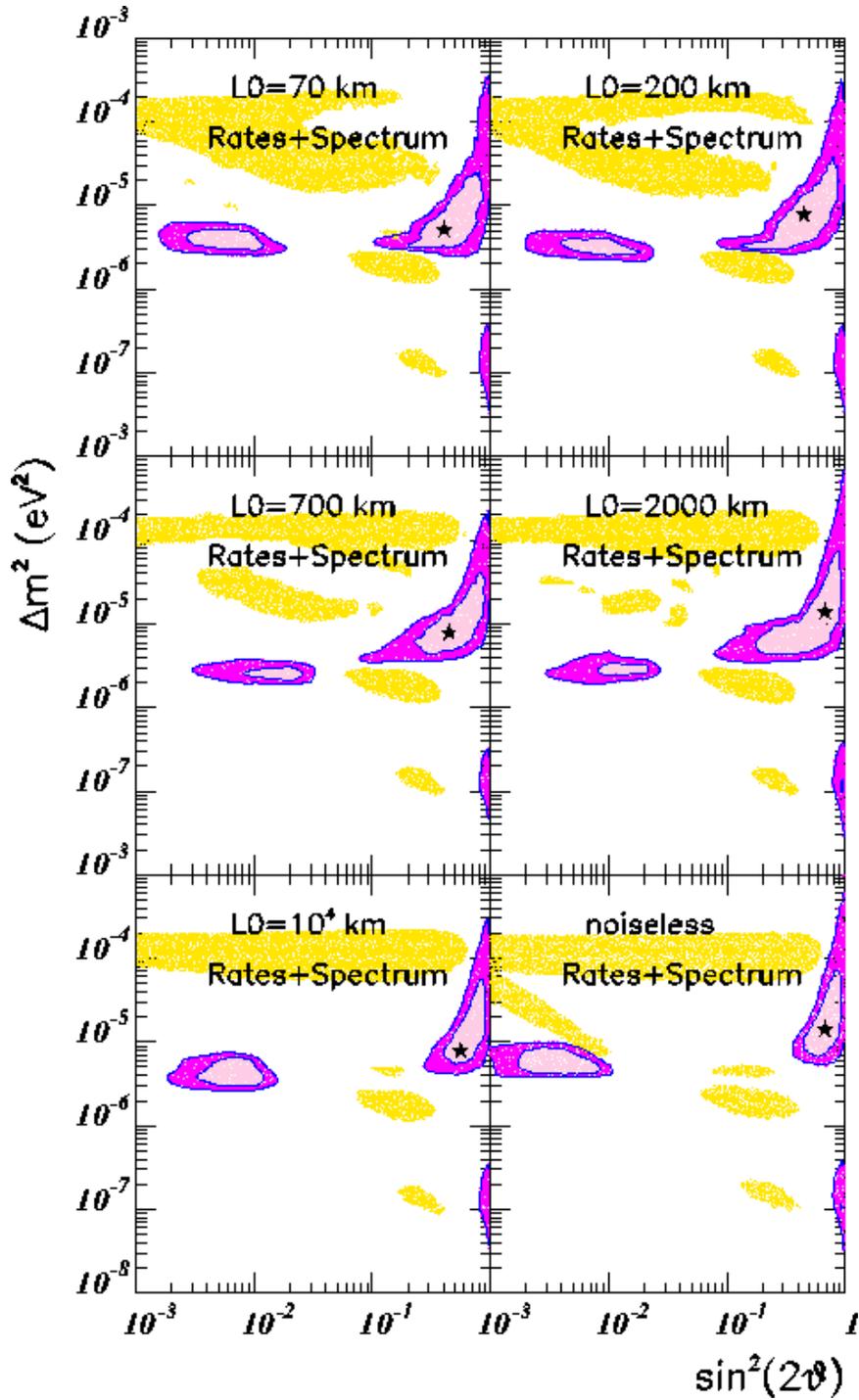,width=0.7\textwidth}}
\end{center} 
\caption{Same as Fig.~\protect\ref{rates} but including also the data on the recoil electron energy spectrum 
observed in Super--Kamiokande. 
The shadowed area represents the region excluded by spectrum  data at 99\% CL.
} 
\label{ratess}
\end{figure}

\begin{figure}
\begin{center}
\mbox{\epsfig{file=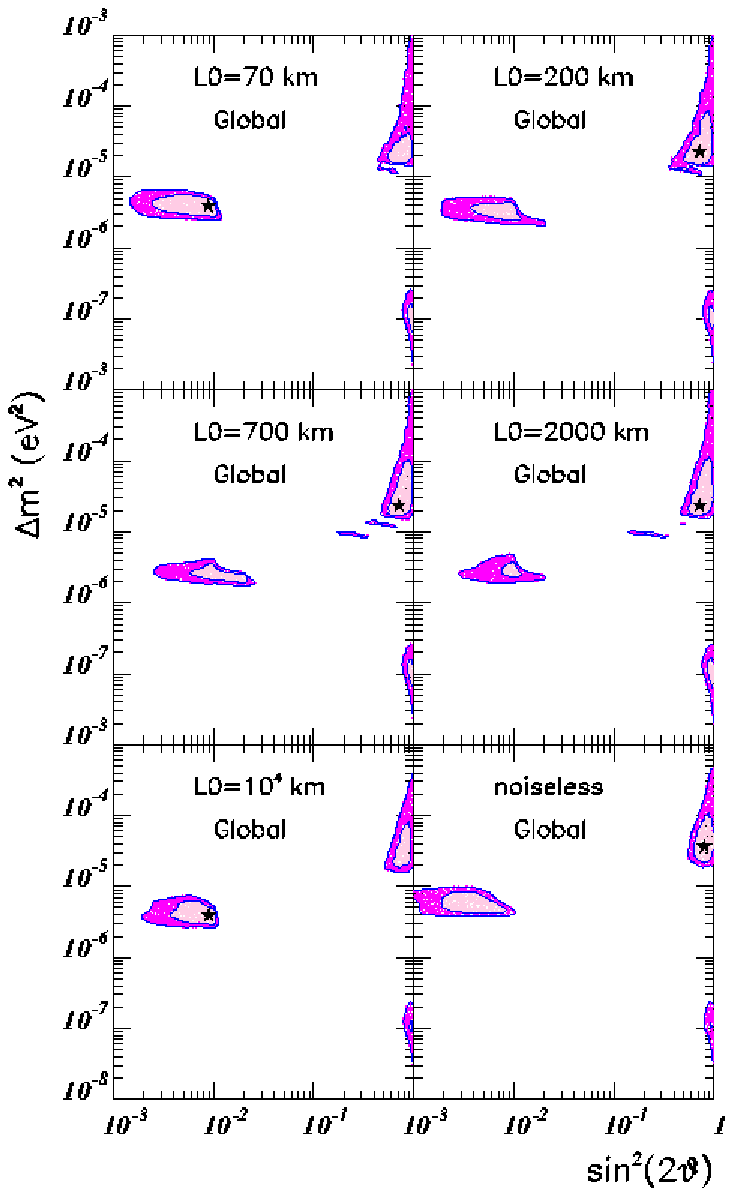,width=0.7\textwidth}}
\end{center}
\caption{Same as Fig.~\protect\ref{rates} for a global fit to the 
solar neutrino data.}
\label{global}
\end{figure}

Next, in order to establish the possible
effect of the presence of noise on the MSW solutions to the SNP we have 
performed a global analysis 
of all the existing observables including not only the measured total rates 
but also the Super--Kamiokande measurement on the time dependence of the 
event rates during  the day and night, as well as the recoil electron energy  
spectrum. The result of such analysis is presented in 
Figs.~\ref{rates}--\ref{global} where we plot the allowed regions for MSW neutrino oscillations
in the framework of  
two--neutrino  mixing with the Sun density profile generated from the BP98, 
after including random noise with amplitude $\xi=4\%$ and different correlation 
lengths $L_{0}$ (70, 200, 700, 2000 y 10000 km). The main conclusions 
are that the total rates are the most sensitive observables to
the presence of noise. On the other hand when the many degrees of 
freedom corresponding to the Super--Kamiokande spectrum are included 
the dependence of the allowed mixing parameters on the matter noise 
is smoothed. This is caused by the larger sensitivity of the mean 
$\langle P_{ee}\rangle$ to the noise for low energy neutrinos. 
Due to its higher energy threshold, the Super--Kamiokande experiment
is mostly insensitive to these effects.
For the same reason one expects that the Borexino experiment 
would be more suitable to place bounds both on the level 
of neutrino noise $\sqrt{<\xi>^2}$ and on the correlation length $L_0$.

\acknowledgments 
This work was supported by DGICYT under grants PB98-0693 and PB97-1261, 
and by the TMR network grant ERBFMRXCT960090 of the European Union.
M.C. Gonzalez-Garcia wish to thank the Phenomenology Institute 
for their kind hospitality during her visit. The work of
A.A. Bykov, V.Yu. Popov and V.B. Semikoz was partially supported through 
RFBR grant 00-02-16271 and for V.B.S. by Iberdrola excellence grant. 


\begin{table} 
\begin{tabular}{|l|l|l|l|l|} 
Experiment & Rate & Ref. & Units& $ R^{\text BP98}_i $\\ 
\hline 
Homestake  & $2.56\pm 0.23 $ & \protect\cite{homestake} & SNU &  $7.8\pm 1.1 $   \\ 
GALLEX + SAGE  & $72.3\pm 5.6 $ & \protect\cite{gallex,sage} & SNU & $130\pm 7 $  \\ 
Kamiokande & $2.80\pm 0.38$ & \protect\cite{kamioka} &  
$10^{6}$~cm$^-2$~s$^{-1}$ & $5.2\pm 0.9 $ \\    
Super--Kamiokande & $2.45\pm 0.08$ & \protect\cite{sk99} &  
$10^{6}$~cm$^{-2}$~s$^{-1}$ & $5.2\pm 0.9 $ \\    
\end{tabular} 
\vglue .3cm 
\caption{Measured rates for the Chlorine, Gallium, Kamiokande and Super--Kamiokande  
experiments. } 
\label{ratesexp} 
\end{table} 
\begin{table}
\begin{tabular}{|c|c|c|c|c|c|c|c|c|c|c|c|c|}
\hline
 Length &\multicolumn{4}{c|}
{$\chi^2_{min,SMA}$}& 
\multicolumn{4}{c|}
{$\chi^2_{min,LMA}$}& 
\multicolumn{4}{c|}
{$\chi^2_{min,LOW}$}\\
\hline
     & Rates & Rates & Rates & Global  
     & Rates & Rates & Rates & Global  
     & Rates & Rates & Rates & Global \\[-0.1cm]
     &       & +Zen & +Spec &      
     &       & +Zen & +Spec &     
     &       & +Zen & +Spec & \\    
\hline
noiseless   &  0.4 & 6.2 &22.3 & 29.2 & 2.9 &7.5  & 22.1   & 27.9 
& 7.4 &12.5  & 26.9 &32.0\\
\hline
$L_{0}=70$ km &  0.2 & 4.6 & 23.4  & 28.4  & 1.9  &7.1  &21.9  & 28.7  
& 7.4 &12.5  & 26.9 & 32.0 \\
\hline
$L_{0}=200$ km &  0.6 &5.5 & 23.7  & 28.9  & 1.5 &7.0  &21.6  &27.6  
& 7.4 &12.5  & 26.9 & 32.0 \\
\hline
$L_{0}=700$ km &  1.5 & 8.0 & 22.6 & 29.8 & 1.7 & 7.4 &21.0  & 27.6  
& 7.4 &12.5  & 26.9 & 32.0 \\
\hline
$L_{0}=2000$ km &  2.8 & 7.5 &25.4  & 30.5  & 2.1 & 7.4 &21.6  &27.6  
& 7.4 &12.5  &26.9  & 32.0 \\
\hline
$L_{0}=10000$ km &  0.2 & 4.8 &22.3   & 27.6  & 2.5 &7.5  & 22.2  & 27.8  
&7.4 &12.5  & 26.9 &32.0\\
\hline
\end{tabular}
\caption{$\chi^2_{min}$ for SMA, LMA and LOW solutions for the total 
event rates.}
\label{chimin}
\end{table}




\begin{thebibliography}{99} 
\bibitem{homestake0} 
R. Davis, Jr, D. S. Harmer, and K. C. Hoffman, Phys. Rev. Lett. {\bf 
20}, 1205 (1968) 
%
\bibitem{SSMold} 
J. N. Bahcall, N. A. Bahcall, and G. Shaviv, Phys. Rev. Lett. {\bf 
20}, 1209 (1968); J. N. Bahcall, R. Davis, Jr., Science {\bf 191}, 264 
(1976). 
%
\bibitem{Bahcall:1997qw} 
J.N.~Bahcall, M.H.~Pinsonneault, S.~Basu and J.~Christensen-Dalsgaard, 
Phys. Rev. Lett. {\bf 78}, 171 (1997) 
 
\bibitem{Bahcall:1995bt} 
J.N.~Bahcall and M.H.~Pinsonneault, Rev. Mod. Phys. {\bf 67}, 781 
(1995) 
 
\bibitem{homestake} 
B.~T.~Cleveland {\em et al.}, Ap.~J. {\bf 496}, 505 (1998).  
 
\bibitem{gallex} 
T.~Kirsten, Talk at the Sixth international workshop on topics in 
astroparticle and underground physics September, TAUP99, Paris, 
September 1999. 
 
\bibitem{sage}  
SAGE Collaboration, J. N. Abdurashitov et al., Phys. Rev. {\bf C60}, 
055801 (1999).
 
\bibitem{kamioka}  
Kamiokande Collaboration, Y. Fukuda et al., Phys. Rev. Lett. {\bf 77}, 
1683 (1996). 
 
\bibitem{sk1}  
Super--Kamiokande Collaboration, Y. Fukuda et al., 
Phys. Rev. Lett. {\bf 82}, 1810 (1999). 
Super--Kamiokande Collaboration, Y. Fukuda et al., 
Phys. Rev. Lett. {\bf 82}, 2430 (1999). 
 
\bibitem{sk99}  
Y. Suzuki, talk a the ``XIX International Symposium on 
Lepton and Photon Interactions at High Energies'', Stanford University,  
August 9-14, 1999;  
M. Nakahata, talk at the  
``6th International Workshop on Topics in Astroparticle and  
Underground Physics, TAUP99'', Paris, September 1999. 

\bibitem{ourseasonal} 
P. C. de Holanda, C. Pe\~na-Garay, M.\ C.\ Gonzalez-Garcia and J.\ W.\ 
F.\ Valle, Phys. Rev. {\bf D60}, 093010 (1999)
\bibitem{v99} 
Talks by M.~C.~Gonzalez-Garcia and A. Yu. Smirnov, Proceedings of {\sl 
International Workshop on Particles in Astrophysics and Cosmology: on 
Theory to Observation} Valencia, May 3-8, 1999, to be published by 
Nucl. Phys. Proc. Supplements, ed. V. Berezinsky, G. Raffelt and 
J. W.~F. Valle (http://flamenco.uv.es//v99.html) 
 
\bibitem{Glashow:1987jj} 
V.N.~Gribov and B.M.~Pontecorvo, Phys. Lett. {\bf 28B}, 493 (1969); 
V. Barger, K. Whisnant, R.J.N. Phillips, Phys. Rev. {\bf D24}, 538 
(1981); S.L.~Glashow and L.M.~Krauss, Phys. Lett. {\bf 190B}, 199 
(1987); V.~Barger, R.J.~Phillips and K.~Whisnant, 
Phys. Rev. Lett. {\bf 65}, 3084 (1990); S.L.~Glashow, P.J.~Kernan and 
L.M.~Krauss, Phys. Lett. {\bf B445}, 412 (1999); V. Berezinsky, 
G.~Fiorentini and M.~Lissia, hep-ph/9811352 and hep-ph/9904225. 
 
\bibitem{msw} 
S.P. Mikheyev and A.Yu. Smirnov, Sov. Jour. Nucl. Phys.  
42, 913 (1985); L.\ Wolfenstein, Phys.\ Rev.\ {\bf D17}, 2369 (1978). 

\bibitem{oursolar} M.C. Gonzalez-Garcia, P.C. de Holanda, C. Pe\~na-Garay and 
J.W.F. Valle, Nucl. Phys. {\bf B573}, 3 (2000). 
.C. Gonzalez-Garcia, C. Pena-Garay, hep-ph/0002186, to appear in  
Phys. Rev. {\bf D}.

\bibitem{Krastev}
P.I. Krastev, A.Yu. Smirnov, Phys. Lett. B226 (1989) 341; Mod. Phys. 
Lett. {\bf A6} (1991) 1001.

\bibitem{otherper}
A. Sch\"afer and S. Kooning, Phys. Lett {\bf B185} (1987) 417.
W. Haxton and W.M. Zhang, Phys. Rev. {\bf D43} (1991) 2484.

\bibitem{other}
F.N. Loreti, A.B. Balantekin, 
Phys. Rev. D50 (1994) 4762; F.N. Loreti, Y.Z. Qian, G.M. Fuller, 
A.B. Balantekin, Phys. Rev. D52 (1995) 6664; E. Torrente-Lujan, 
hep-ph/9602398; A.B. Balantekin, F.N. Loreti, Phys. Rev. D54 (1996) 3941;

\bibitem{Burgess0}
C.P. Burgess, D. Michaud, Ann. Phys. NY, {\bf 256} (1997) 1.

\bibitem{Nunokawa}
H. Nunokawa, A. Rossi, V.B. Semikoz, J.W.F. Valle, Nucl. Phys. B472 
(1996) 495.

\bibitem{Burgess} 
P. Bamert, C.P. Burgess, D. Michaud, Nucl. Phys. {\bf B513} (1998)319.

\bibitem{Dzhalilov} 
N.S. Dzhalilov, V.B. Semikoz, astro-ph/9812149. V.B. Semikoz and 
N.S. Dzhalilov, {\it Xth International School "PARTICLES and COSMOLOGY"}, 
Baksan Valley, Kabardino-Balkaria, Russian Federation. PP.101-109 (1999)

\bibitem{daynight} 
J. Bouchez {\it et. al.}, Z. Phys. {\bf C32}, 499 (1986);
S. P. Mikheyev and A. Yu. Smirnov, {\it '86 Massive Neutrinos in
Astrophysics and in Particle Physics}, proceedings of the Sixth
Moriond Workshop, edited by O. Fackler and J. Tr$\hat{a}$n Thanh
V$\hat{a}$n (Editions Fronti\`eres, Gif-sur-Yvette, 1986), pp. 355;
S.P. Mikheyev and A.Yu. Smirnov, Sov. Phys. Usp. 30 (1987) 759-790;
A. Dar {\it et. al.} Phys. Rev. {\bf D 35} (1987) 3607;
 E. D. Carlson,Phys. Rev. {\bf D34}, 1454 (1986) ; 
A.J. Baltz and J. Weneser,Phys. Rev. {\bf D50}, 5971 (1994);
 A. J. Baltz and J. Weneser,Phys. Rev. {\bf D51}, 3960 (1994); 
P. I. Krastev, hep-ph/9610339;
Q.Y. Liu, M. Maris and S.T. Petcov, Phys. Rev. {\bf D56}, 5991 (1997);
M. Maris and S.T. Petcov, Phys. Rev. {\bf D56}, 7444 (1997);
J.N. Bahcall and P.I. Krastev, Phys. Rev. {\bf C56}, 2839 (1997);
A. J. Baltz and J. Weneser, Phys. Rev. {\bf D35}, 528 (1987);
A. J. Baltz and J. Weneser, Phys. Rev. {\bf D37}, 3364 (1988);
E.~Lisi and D.~Montanino, Phys. Rev. {\bf D56}, 1792 (1997);
S.~T. Petcov, Phys. Lett. {\bf B434}, 321 (1998);
M.~Chizhov, M.~Maris, and S.~T. Petcov, (1998), hep-ph/9810501; 
M.V. Chizhov and S.T. Petcov, Phys. Rev. Lett. {\bf 83}, 1096 (1999);
A.S. Dighe, Q.Y. Liu and A.Yu. Smirnov, hep-ph/9903329; 
A.H. Guth, L. Randall and M. Serna, J. High Energy Phys. {\bf 8}, 018 (1999).


\bibitem{Dziembowski}
S.S. Degl'Innocenti, W.A. Dziembowski, G. Fiorentini, B. Ricci, Astropart. 
Phys. 7 (1997), 77.

\bibitem{Dziembowski1}
W.A. Dziembowski, Bull. Astron.Soc.India 24 (1996) 133;\\also there reviews:
Jorgen Christensen-Dalsgaard, Lecture Notes, available at http://bigcat.
obs.aau.dk/ jcd/oscilnotes/.\\S. Turck-Chieze et al., Phys. Rep. 230 (1993) 57.

\bibitem{LZ} 
See, for instance, 
P.I.~Krastev and S.T.Petcov, Phys. Lett. {\bf B207}, 64 (1988);
S.T.Petcov, Phys. Lett. {\bf B200}, 373 (1988) and references 
therein. 

\bibitem{Chechin}
V.K. Ermilova, V.A. Tsarev and V.A. Chechin, kr. Soob. Fiz. 
Lebedev Institute 5 (1986) 26. 


\bibitem{lisi3} G.L. Fogli, E. Lisi, D. Montanino and A. Palazzo,  
hep-ph/9912231.

\bibitem{BP98}  
J.N. Bahcall, S. Basu and M. Pinsonneault, Phys. Lett. B433 (1998) 1. 
 
\bibitem{prod}  
http://www.sns.ias.edu/\~{}jnb/SNdata 
 
\bibitem{Faid}  
B.\ Fa\"{\i}d, G.\ L.\ Fogli, E.\ Lisi and D.\ Montanino, 
Phys.\ Rev.\ {\bf D55}, 1353 (1997). 
%
\bibitem{CrSe} 
J. N. Bahcall, M.\ Kamionkowsky, and A.\ Sirlin, Phys.\ Rev.\ {\bf 
D51}, 6146 (1995). 
 
\bibitem{fogli-lisi}  
G.\ L.\ Fogli, E.\ Lisi and D.\ Montanino, Phys.\ Rev.\ D {\bf 49}, 
3226 (1994). G.\ L.\ Fogli, E.\ Lisi, Astropart. Phys.{\bf 3}, 185 
(1995). 
 
\end{thebibliography}
\end{document}